\documentclass[twocolumn,prb,superscriptaddress]{revtex4}
\usepackage{graphicx}
\usepackage{epsfig}

\def\beq{\begin{equation}}
\def\eeq{\end{equation}}
\begin{document}
\title{Charge states of strongly correlated 3d oxides: from typical insulator to unconventional electron-hole Bose liquid}
\author{A.S. Moskvin}
\affiliation{Department of Theoretical Physics, Ural State University,  620083,
Ekaterinburg, Russia}

\begin{abstract}
We develop a model approach to describe  charge fluctuations and different charge phases in strongly correlated 3d oxides.  
As a generic model system one considers that of centers each with three possible valence states $M^{0,\pm}$ 
described in frames of $S=1$ pseudo-spin (isospin) formalism by an effective anisotropic non-Heisenberg Hamiltonian which includes two types of single particle correlated hopping and the two-particle hopping. We show that the coherent states provide the optimal way both to a correct mean-field approximation  and respective continuous models to describe the pseudo-spin system. Simple uniform 
mean-field   phases include an insulating monovalent $M^0$-phase, mixed-valence binary (disproportionated) $M^{\pm}$-phase, and mixed-valence ternary (``under-disproportionated'') $M^{0,\pm}$-phase. We consider two first  phases in more details focusing on  the  problem of electron/hole states and different types of excitons in $M^0$-phase and formation of electron-hole Bose liquid in $M^{\pm}$-phase. 
Pseudo-spin formalism provides a useful framework for  revealing and describing different  topological charge fluctuations, in particular, like domain walls or bubble domains in antiferromagnets. 
Electron-lattice polarization effects are shown to be crucial for the stabilization of either phase. All the insulating systems such as $M^0$-phase may be subdivided to two classes: stable and unstable ones with regard to the formation of self-trapped charge transfer (CT) excitons. The latter systems appear to be unstable with regard to the formation of CT exciton clusters, or droplets of the electron-hole Bose liquid.  
The model approach suggested is believed to be applied to describe a physics of
 strongly correlated oxides such as cuprates, manganites, bismuthates, and other systems with
 charge transfer excitonic instability and/or mixed valence.  We shortly discuss an unconventional scenario
of the essential physics of cuprates and manganites that implies their instability with regard to the self-trapping of charge transfer excitons and the formation of electron-hole Bose liquid. 
\end{abstract}
%E-mail: alexandr.moskvin@usu.ru
%\begin{keyword}
% mixed-valence, electron correlations, oxides, cuprate superconductors,
 %pseudo-spins
%\end{keyword}
%\end{frontmatter}
%\pacs{71.10.-w, 71.28.+d, 71.30.+h, 74.72.-h}

%71.10.-w Theories and models of many-electron systems
%
%71.28.+d Narrow-band systems; intermediate-valence solids
%
%71.30.+h Metal-insulator transitions and other electronic transitions
%
%74.72.-h Cuprate superconductors - high-T_c and insulating parent compounds
%

%\keywords{mixed-valence, electron correlations, pseudo-spins, oxides, cuprate
%superconductors, manganites}

\maketitle

\section{Introduction}

The discovery of the high-$T_c$ superconductivity in doped
cuprates,\cite{Muller} observation of many unconventional properties in doped
manganites with their colossal magnetoresistance, bismuthates with
high-$T_c$'s, nickellates and many other oxides \cite{Imada} shows that we deal
with a manifestation of novel strongly correlated states with a local charge
instability, mixed valence, "metal-dielectric"  duality, strong coupling of
different (charge, spin, orbital, structural) degrees of freedom and non-Landau
behaviour of quasiparticles. All this has generated a flurry of ideas,
 models and scenarios of the puzzling transport phenomena   and
 stimulated the intensive studies of various correlation effects and charge transfer (CT) phenomena
 in strongly correlated systems derived in either way from insulators unstable
  with regard to the CT fluctuations.
Conventional approach to hotly debated strongly correlated 3d oxides such as cuprates, manganites, and many other similar systems  implies making use of  a Hubbard model with famous Hamiltonian 
 
\begin{equation}
\hat H = -\sum _{<i,j>,\sigma} t(ij){\hat c}^{\dag}_{i\sigma}{\hat
c}_{j\sigma} +U\sum _{i,\sigma \sigma ^{\prime}}n_{i\sigma}n_{i\sigma
^{\prime}}	\label{Hubbard}
\end{equation}
with competing contributions of kinetic and potential terms. Here ${\hat c}^{\dag}_{i\sigma}/{\hat
c}_{j\sigma}$ are creation/annihilation operators for low-lying antibonding 3d-O 2p hybridized orbitals. The two-center charge transfer integral $t(ij)$ is often associated with d-d transfer. Mott-Hubbard insulator is believed to arise from a potentially metallic half-filled band as a result of the Coulomb blockade of electron tunnelling ($U\gg t$) to neighboring sites.\cite{Mott}

Despite intense effort, the behavior  of strongly correlated 3d oxides remain poorly understood and we are still far from a comprehensive understanding of the underlying physics.  Moreover, it seems that there are  missing qualitative aspects of the problem beyond the simple Hubbard scenario that so far escaped the identification and the recognition. Firstly it concerns strong electron-lattice polarization effects  which may be subdivided into
 electron-lattice interaction itself,\cite{Shluger,Vikhnin} and a contribution of an electronic background that is electronic subsystem which is not incorporated into effective Hubbard model  Hamiltonian.\cite{Hirsch,shift} 
 These effects are of great importance for the
ground state electronic and crystalline structure, and  can seriously modify
the doping response of 3d oxide up to the crucial change of the seemingly
natural ground state. This question has not received the attention it deserves. It should be emphasized that traditional  Fr\"{o}hlich approach to the
electron-lattice coupling implies  the description of linear effects whereas
the charge fluctuations  in the insulator do imply strongly nonlinear
electron-lattice coupling with the predominance of polarization and relaxation
effects, and another energy scale.

 Electron-lattice effects may be directly incorporated into effective Hubbard model. Assuming the coupling with the  local displacement (configuration) coordinate $Q$  in the effective potential energy  we
arrive at a generalized Peierls-Hubbard model.\cite{Nasu} From the other hand, the
taking account of similar effects in the  kinetic energy results in a generalized  Su-Schrieffer-Heeger (SSH)
model.\cite{SSH} The correlation effect of an electronic background was shown \cite{Hirsch,shift} to be of primary importance for atomic systems with filled or almost filled electron shells. Namely such a situation is realized in oxides with O$^{2-}($2p$^6$) oxygen ions. In particular, the effect results in a correlated character of a charge transfer that seems to be one of the main features for 3d oxides.

Many  strongly correlated 3d oxides reveal anomalous sensitivity to a small  nonisovalent substitution. For example, only 2\% Sr$^{2+}$ substituted for La$^{3+}$ in La$_2$CuO$_4$  result in a dramatical suppression of long-range copper antiferromagnetism, while it is suppressed with isovalent Cu$^{2+}$ substitution by Zn$^{2+}$ at a much higher concentration close to the site dilution percolating threshold. Simultaneously, the transport properties of La$_{2-x}$Sr$_x$CuO$_{4}$ system reveal unconventional insulator-metal duality starting from very low dopant level.\cite{Ando}  Most likely, all this points to a charge phase instability intrinsic for parent 214 system which somehow evolves with nonisovalent substitution due to a well developed charge potential inhomogeneity and/or hole doping effect. The problem seems to be closely related with the hidden $multistability$ intrinsic to each
solid.\cite{Toyozawa,Nasu} If the ground state of a solid is
pseudo-degenerate, being composed of true and false ground states with each
structural and electronic orders different from others, one might call it {\it
multi-stable}. Below we focus ourselves on a charge degree of freedom and
charge (in)stability, rather than orbital or spin degrees of freedom. As an
illuminating example of such a material with {\it conceptually simple} but {\it
actually false} ground state Toyozawa \cite{Toyozawa} suggests to address the
Wolfram's red, a quasi-one dimensional material of which the skeleton chain
consists of alternate array: (Cl$^-$ - Pt$^{3+}$ -)$^{2n}$ with simple (and
seemingly metallic), but a false ground state. The real ground state is an
insulator with a complicated structure of doubled period: (Cl$^{-}$ - Pt$^{4+}$
- Cl$^{-}$ - Pt$^{2+}$-)$^n$, which  can be reached from the former through the
Peierls transition with the charge  density wave of large amplitude, or
disproportionation like reaction. This  transition can be considered as the
condensation of self-decomposed  self-trapped excitons spontaneously generated
on all unit cells.

In this connection it is worth noting the text-book example of BaBiO$_3$ system
where we unexpectedly deal with the disproportionated Ba$^{3+}$+ Ba$^{5+}$
ground state instead of the conventional lattice of Ba$^{4+}$
cations.\cite{Kagan} The bismuthate situation can be viewed also as a result of
a condensation of CT excitons, in other words, the spontaneous
generation of self-trapped CT excitons in the ground state
with a proper transformation of lattice parameters.
 At present, a CT instability  with regard to disproportionation is believed to be a rather
typical property for a number of perovskite 3d transition-metal oxides
 such as SrFeO$_3$, LaCuO$_3$, RNiO$_3$ \cite{Mizokawa}, moreover,  in solid state chemistry  one consider tens of
disproportionated systems.\cite{Ionov} 
New principles must be developed to treat  such charge or CT unstable systems with their dramatical non-Fermi-liquid behavior. In particular, we have to change the current paradigm
of the metal-to-insulator (MI) transition to that of an insulator-to-metal (IM)
phase transition. These two approaches imply essentially different starting
points: the former starts from a rather simple {\it metallic-like} scenario with inclusion of
correlation effects, while the latter does from strongly correlated {\it atomic-like}
scenario with the inclusion of a charge transfer. Electron-lattice polarization effects accompanying the charge transfer appear to be of primary importance to stabilize either phase state.  One should emphasize that
the theoretical description of such systems is one of the challenging problems in solid state physics.
 
Hereafter, we develop a model approach to describe different charge fluctuations and charge phases in strongly correlated 3d oxides with main focus on the correlated CT effects. As an illustrative model system we address a simple  mixed-valence system with three possible stable nondegenerate valent states of a cation-anionic cluster, hereafter $M$: $M^0, M^{\pm}$, forming the charge (isospin) triplet. The $M^0$ valent  state is associated
 with the {\it conceptually simple} one like CuO$_{4}^{6-}$ in insulating
 copper oxides (CuO, La$_2$CuO$_4$, YBa$_2$Cu$_3$O$_6$, Sr$_2$CuO$_2$Cl$_2$,...)
 or MnO$_{6}^{9-}$ in manganite LaMnO$_3$ or BiO$_{6}^{9-}$ in bismuthates. It is worth noting  that such a model is a most relevant to describe different cuprates where novel concepts should compete with a
 traditional  Hubbard model approach in a hole representation implying the vacuum state formed by $M^-$ (CuO$_{4}^{5-}$) centers,  and some concentration of holes. That is why overall the paper we refer the insulating cuprates to illustrate the main concepts of the approach developed.  Our mathematics is based on the $S=1$ pseudo-spin formalism (see e.g. review article Ref.\onlinecite{pseudo}) to be the effective tool for the
description of the  essential physics both of insulators unstable with regard to the CT fluctuations and related mixed-valence systems. Such an approach provides the  universal framework for a unified
  description of these systems as possible phase states of a certain {\it parent
  multi-stable system}. In addition, we may make use of  powerful methods developed in
  the physics of spin systems. The model system  of $M^{0,\pm}$ centers  is 
described in frames of $S=1$ pseudo-spin formalism by an effective anisotropic non-Heisenberg Hamiltonian which includes two types of correlated two-center hopping:
$$
M^0 +M^0 \leftrightarrow M^{\pm}+M^{\mp}\, \mbox{and} \,M^{\pm} + M^0 \leftrightarrow M^0 +M^{\pm},
$$
respectively. It should be noted that we neglect all the intra-center transition, including anion-cation O 2p-3d charge transfer.

Our main goal  is to describe different charge phases of the model system and a scenario of evolution of visibly typical insulator to unconventional electron-hole Bose liquid which reveals many unexpected  properties, including superconductivity.
The paper is organized as follows: In Sec.II we address a metal-oxide cluster model, different mechanisms of correlation effects, and the effects of electron-lattice polarization. In Sec.III we introduce the $S=1$
pseudo-spin formalism to describe the model mixed-valence systems. The effective pseudo-spin 
Hamiltonian and possible mean-field phase states of the mixed-valence systems are discussed in Sec.IV. In Sec.V we analyse an eh-representation of different excitations in a monovalent $M^0$ phase, discuss  a CT instability, and nucleation of electroh-hole (EH) droplets. Electron-hole Bose liquid is discussed in Sec.VI. Some topological skyrmion-like charge
fluctuations in the model mixed valence system are described in Sec.VII.  Implications for cuprates and manganites
are discussed in Sec.VIII.

\section{Metal-oxide clusters and correlation effects}
The electronic states in  strongly correlated 3d oxides manifest
both significant correlations and dispersional features. The
dilemma posed by such a combination is the overwhelming number of
configurations which must be considered in treating strong
correlations in a truly bulk system. One strategy to deal with
this dilemma is to restrict oneself to small 3d-metal-oxygen clusters, creating
model Hamiltonians whose spectra may reasonably well represent the
energy and dispersion of the important excitations of the full
problem. Indeed, such clusters as CuO$_4$ in quasi-2D cuprates, MnO$_6$ in manganite perovskites are basic elements of crystalline and electronic structure.
Despite a number of principal
shortcomings, including the boundary conditions, the breaking of
local symmetry of boundary atoms, sharing of common anions for $nn$ clusters etc., the 
embedded molecular cluster method provides both, a clear physical
picture of the complex electronic structure and the energy
spectrum, as well as the possibility of quantitative modelling.
Eskes {\it et al.} \cite{Eskes}, as well as Ghijsen {\it et al.}
\cite{Ghijsen} have shown that in a certain sense the cluster
calculations might provide a better description of the overall
electronic structure of  insulating  3d oxides  than
band-structure calculations.  In particular, they allow to take better into
account different correlation effects.

Below, in the Section we discuss some aspects of electronic structure, energy spectrum, and correlation effects for an illustrative example of CuO$_4$ clusters embedded into an insulating cuprate. 

\subsection{Electronic structure of copper-oxygen clusters}

Beginning from 5 Cu $3d$ and 12 O $2p$ atomic orbitals for CuO$_4$
 cluster with $D_{4h}$ symmetry, it is easy to form 17 symmetrized
 $a_{1g},a_{2g},b_{1g},b_{2g},e_{g}$ (gerade=even) and $a_{2u},b_{2u},
 e_{u}(\sigma),e_{u}(\pi)$
 (ungerade=odd) orbitals. The even Cu $3d$ $a_{1g}(3d_{z^2}), b_{1g}(3d_{x^2-y^2}),
 b_{2g}(3d_{xy}), e_{g}(3d_{xz},3d_{yz})$
 orbitals hybridize, due to strong Cu $3d$-O $2p$ covalency, with even O
$2p$-orbitals
 of  the same symmetry, thus forming appropriate bonding
$\gamma ^{b}$ and antibonding $\gamma ^{a}$ states. Among the odd
orbitals only $e_{u}(\sigma)$ and $e_{u}(\pi)$ hybridize due to
nearest neighbor $pp$ overlap and transfer thus forming appropriate
bonding $e_{u}^{b}$ and antibonding $e_{u}^{a}$ purely oxygen
states. The purely oxygen $a_{2g},a_{2u},b_{2u}$ orbitals are
nonbonding.
All "planar" O $2p$ orbitals  in accordance
with the orientation of lobes could be classified as $\sigma$
($a_{1g},b_{1g},e_{u}(\sigma)$) or $\pi$
($a_{2g},b_{2g},e_{u}(\pi)$) orbitals, respectively.

Bonding and antibonding molecular orbitals in hole representation
can be presented, for example, as follows
$$
|b_{1g}^{b}\rangle =\cos \alpha _{b_{1g}} |b_{1g}(3d)\rangle + \sin
\alpha _{b_{1g}}|b_{1g} (2p)\rangle ,
$$
\begin{equation}
| b_{1g}^{a}\rangle =\sin \alpha _{b_{1g}} |b_{1g}(3d)\rangle - \cos
\alpha _{b_{1g}}|b_{1g}(2p)\rangle ,
\end{equation}
where $b_{1g}(3d)=3d_{x^2-y^2}$ and $|b_{1g} (2p)\rangle$ is a superposition of four O 2p orbitals with $b_{1g}$ symmetry.

\begin{figure}[t]
\includegraphics[width=8.5cm,angle=0]{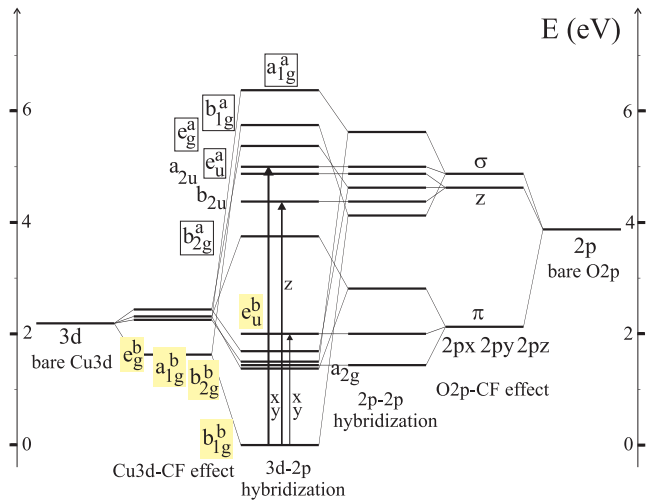}
\caption{Model single-hole energy spectra for a CuO$_4$ plaquette
with parameters relevant for  a number of
 insulating cuprates.} \label{fig1}
\end{figure}
Fig. \ref{fig1}  presents a  single-hole energy spectrum
for a CuO$_4$ plaquette embedded into an insulating cuprate like
Sr$_2$CuO$_2$Cl$_2$ calculated with a reasonable set of parameters.\cite{CT} For
illustration we show also a  step-by-step formation of the cluster
energy levels from the bare Cu $3d$ and O $2p$ levels with the
successive inclusion of crystalline field (CF) effects, O $2p$-O
$2p$, and Cu $3d$-O $2p$ covalency. It is worth noting that strong Cu 3d-O 2p overlap and covalency result in a dramatic difference between CuO$_4^{6-}$ center and Cu$^{2+}$ cation as its naive analogue. 

\subsubsection{Two-hole configurations for the CuO$_{4}^{5-}$
center}

Starting from CuO$_{4}^{6-}$ center as a realization of $M^0$ center we arrive at CuO$_{4}^{5-}$ and CuO$_{4}^{7-}$
centers as realizations of  hole $M^+$ and  electron $M^-$ centers, respectively. The electron  CuO$_{4}^{7-}$
center has a filled set of Cu 3d and O 2p orbitals, and is often addressed to be a generalization of Cu$^{1+}$ ion. 
The hole  CuO$_{4}^{5-}$ center has actually two-hole configuration with  the lowest Zhang-Rice (ZR) spin-singlet $^1A_{1g}$ state \cite{ZR} formed by the interaction of three  "covalent" configurations: $(b_{1g}^a)$, $(b_{1g}^b)$, and $(b_{1g}^ab_{1g}^b)$, respectively, or three "purely ionic" two-hole configurations
$|d^{2}\rangle$, $|pd\rangle$, and $|p^{2}\rangle$. Here, $|d\rangle=|d_{x^2 -y^2}\rangle$ and
$|p\rangle=|p_{b_{1g}}\rangle$ are the non-hybridized Cu 3$d_{x^2-y^2}$ and O 2$p_{\sigma}$ orbitals, respectively, with bare energies $\epsilon_{d}$ and $\epsilon_{p}$. 

Let us present a simple example of the calculation of the
two-hole spectrum in the ZR-singlet sector. 
 The matrix of
the full effective Hamiltonian within the bare purely ionic basis set has  a
rather simple form
\begin{equation}
{\hat H} =\pmatrix{2\epsilon _{d}+U_{d} & t & 0 \cr t & \epsilon
_{d}+\epsilon _{p}+V_{pd} & t \cr 0 & t & 2\epsilon _{p}
+U^{*}_{p}\cr}, \label{U}
\end{equation}
where the effective Coulomb parameter for purely oxygen
configuration incorporates both the intra-atomic parameter $U_p$
and the oxygen-oxygen coupling to the first and second nearest
neighbors, respectively
$$
U^{*}_{p}=U_{p}+ \frac{1}{4}V_{pp}^{(1)}+\frac{1}{8}V_{pp}^{(2)},
$$
and the following condition holds:
$$ U_d \, > \,U_p \, >\, V_{pd}
$$
For reasonable values of parameters (in eV): $U_{d}= 8.5$,
$U_{p}=4.0$, $V_{pd}=1.2$, $\epsilon _{d}=0$, $\epsilon _{p}=3.0$,
$t=t_{pd}=1.3$ (see Ref.\  \onlinecite{CT}) we obtain for
the ZR-singlet energy $E_{ZR}=3.6$, and its wave function
\begin{equation}
|\Phi_1^{(2)}\rangle=|b_{1g}^{2};pd\rangle = 0.25|d^{2}\rangle
-0.95|pd\rangle  +0.19|p^{2}\rangle \; , \label{zr1}
\end{equation}
where $|pd\rangle=\frac{1}{\sqrt{2}}(|p(1)d(2)\rangle +|p(2)d(1)\rangle)$.
It reflects the well-known result that the ZR-singlet represents a
two-hole configuration with one predominantly Cu $3d$ and one
predominantly O $2p$ hole. It is worth noting that the
hole $CuO_4^{5-}$ center sometimes one  naively associate  with $Cu^{3+}$ ion, however,  such a conclusion is a rather far from reality. Indeed, this  center is  the mixed valence one, as  the $Cu$
valence resonates between $+2$ and $+3$.

The two excited states with energies
$E_{ZR}+5.2$ and $E_{ZR}+6.7$ eV are described by the wave
functions
\begin{equation}
|\Phi_2^{(2)}\rangle=|b_{1g}^{2};dd\rangle = -0.95|d^{2}\rangle
-0.21|pd\rangle +0.22|p^{2}\rangle , \label{zr2}
\end{equation}
\begin{equation}
|\Phi_3^{(2)}\rangle=|b_{1g}^{2};pp\rangle = 0.17|d^{2}\rangle
+0.24|pd\rangle +0.96|p^{2}\rangle , \label{zr3}
\end{equation}
respectively. Given the ZR-singlet energy one may calculate the minimal
 energy for $M^0 +M^0 \rightarrow M^{\pm}+M^{\mp}$ charge transfer:
$$
\Delta_{CT}=E_{ZR}-2E_{b_{1g}}=(3.6 +0.5) \mbox{eV} = 4.1
\mbox{eV},
$$
where the stabilization energy for the bonding $b_{1g}^b$ state is
simply calculated from matrix (\ref{U}) at
$U_{d}=U^{*}_{p}=V_{pd}=0$.  It should be emphasized that this quantity plays a
particular role as the minimal charge transfer energy which
specifies the charge transfer gap.  In the general case it is
defined as follows:
$$
\Delta _{CT}= E_{N+1}+E_{N-1}-2E_{N},
$$
or as the energy required to remove a hole from one region of the
crystal and add it to another region beyond the range of excitonic
correlations. The exact diagonalization studies for a series of
clusters with different size \cite{Hybertsen,Hybertsen1} show that
$\Delta _{CT}$ strongly diminishes with cluster size from $\approx
4$eV for small clusters to $\approx 2.5$eV as extrapolated value
for large clusters.

Our simple calculation points to a significant role of correlation effect. Indeed, the inter-configurational coupling  due to Coulomb repulsion results in a visible deviation of the two-hole ground state  wave function from the predictions of simple model of noninteracting particles
$$
|(b_{1g}^b)^{2}\rangle = (-0.8|d(1)\rangle
+0.6|p(1)\rangle )  (-0.8|d(2)\rangle
+0.6|p(2)\rangle )
$$
\begin{equation}
=0.64|d^{2}\rangle
-0.68|pd\rangle  +0.36|p^{2}\rangle \,.
\label{12}
\end{equation}
 In particular, it could give rise to a strong renormalization of the hole transfer integrals. 

\subsection{Electron-lattice polarization effects}

\subsubsection{Correlation effects of electronic background}
The correlation problem becomes of primary importance for atoms/ions near Coulomb instability when the one-electron gluing cannot get over the destructive effect of the electron-electron repulsion. Such a situation seems to realize in oxides where Hirsch {\it et al}. \cite{Hirsch} have proposed an instability of  O$^{2-}$(2p$^6$) electronic background. 
 The main suggestion in their theory of "anionic metal" concerns
the occurrence of the non-rigid degenerate structure for a closed
electron shell such as O$^{2-}$(2p$^6$)   with the internal purely
correlation degrees of freedom. In other words, one should expect
sizeable correlation effects not only from unfilled 3d- or oxygen 2p
shells, but from completely filled O 2p$^6$ shell! In order to
relevantly describe such a non-rigid atomic background and its
coupling to the valent hole one might use a concept of the
well-known "shell-droplet" model for nuclei after Bohr and
Mottelson.\cite{Nataf} In accordance with the model a set of
completely filled electron shells which form an atomic background or
vacuum state for a hole representation is described by certain
internal collective degrees of freedom and a number of physical
quantities such as electric quadrupole and magnetic moments. Valent
hole(s) moves around this non-rigid background with strong
interaction inbetween. Such an approach strongly differs from the
textbook one that implies a rigid atomic orbital basis irrespective
of varying filling number and external potential.

 None of the effective many-body Hamiltonians that are
most widely used to study the effect of electron correlation in
solids such as the Hubbard model, the Anderson impurity and lattice
models, the Kondo model, contain this very basic and fundamental
aspect of electron correlation that follows from the atomic
analysis. \cite{Hirsch1} The Hubbard on-site repulsion U between
opposite spin electrons on the same atomic orbital is widely
regarded to be the only important source of electron correlation in
solids. It is a clear oversimplification, and we need in a more
realistic atomic models to describe these effects, especially for
atoms in a specific external potential giving rise to a Coulomb
instability. To this end  we have proposed a generalized non-rigid shell model (see Refs.\onlinecite{shift,shift1,shift2}). The model
represents a variational method for the many-electron atomic
configurations with the trial parameters being the coordinates of
the center of the one-particle atomic orbital.
The resulting displacement of the atomic orbitals allows a simple
interpretation of the electron density redistribution stemmed from taking into account the electron-electron repulsion, and the
symmetry of a system can be readily used for the construction of the
trial many-electron wave function.  As a whole, the model bears a
strong resemblance to the conventional well-known shell model by
Dick and Overhauser \cite{shell} widely used in lattice dynamics. In
frames of the model the ionic configuration with filled electron
shells is considered to be constituted of an outer spherical shell
of 2(2l+1) electrons and a core consisting of the nucleus and the
remaining electrons. In an electric field the rigid shell retains
its spherical charge distribution but moves bodily with respect to
the core. The polarizability is made finite by a harmonic restoring
force of spring constant $k$ which acts between the core and shell.
The shells of two ions repel one another and tend to become
displaced with respect to the ion cores because of this repulsion.
The respective displacement vector appears to be a simplest
$collective$ coordinate which specifies the change of the
electron-nucleus attraction. It should be noted that such a
displacement does not imply any variation in electron-electron
repulsion and respective correlation energy.

However, as we shall see below, a simple shell model can be easily
generalized to take account of correlation effects. To this end we
must consider the displacements of separate one-electron orbitals to
form the set of the variational parameters in a correlation
function. Then we can introduce both the displacement of the center
of "gravity" for filled shell and a set of the relative
displacements of separate one-electron orbitals with regard to each
other. The former form an "acoustical" mode and are described in
frames of conventional shell model, while the latter form different
novel "optical" modes. Such a seemingly naive non-rigid shell
picture can provide both the microscopic substantiation of the
conventional shell model and its generalization. Moreover, this {\it
non-rigid shell model} points to a physically clear procedure to
account for the correlation effects. Indeed, the "optical"
displacement mode is believed to provide the minimal
electron-electron repulsion. The non-rigid shell represents a novel
specific atomic state that can be remarkably realized near a Coulomb
instability.

The idea of displaced atomic shells has appeared many years ago
\cite{Kozman}  in the very beginning of the quantum chemical era,
and reflected the naive picture of the repelling electrons. However,
the physically sound idea did not receive the relevant position in
the hierarchy of correlation effects.

\subsubsection{Electron-lattice relaxation effects}
As it is mentioned above, the minimal energy cost of the optically excited
disproportionation or electron-hole formation in insulating cuprates is
$2.0-2.5$ eV. However, the question arises, what is the energy cost for the thermal
excitation of such a local disproportionation?  The answer implies
first of all the knowledge of relaxation energy, or  the energy gain due to the lattice polarization by the
localized charges. The full polarization energy $R$  includes the cumulative
effect of $electronic$ and $ionic$ terms, associated with the displacement of
electron shells and ionic cores, respectively.\cite{Shluger} The former term $R_{opt}$ is due
to the {\it non-retarded} effect of the electronic polarization by the momentarily
localized electron-hole pair given the ionic cores fixed at their perfect crystal positions.
Such a situation is typical for lattice response accompanying the Franck-Condon
transitions (optical excitation, photoionization). On the other hand, all the
long-lived excitations, i.e., all the intrinsic thermally activated states and
the extrinsic particles produced as a result of doping, injection or optical
pumping should be regarded as stationary states of a system with a deformed
lattice structure. These relaxed states should be determined from the condition
that the system energy has a local minimum when account is taken of the
interaction of the electrons and holes with the lattice deformations. At least,
it means that  we cannot, strictly speaking, make use of the same energy
parameters to describe the optical (e.g. photoexcited) hole and thermal (e.g.
doped) hole.

For the illustration of polarization effects in cuprates we apply the shell
model calculations to look specifically at energies associated with the
localized holes of Cu$^{3+}$ and O$^-$ in "parent" La$_2$CuO$_4$ compound.
It follows from these calculations that there is a large difference in the
lattice relaxation energies for O$^-$ and Cu$^{3+}$ holes. The lattice
relaxation energy, -$\Delta R^{\alpha}_{th}$, caused by the hole localization
at the O-site (4.44 eV) appears to be significantly larger than that for the
hole localized at the Cu-site (2.20 eV). This indicates
the strong electron-lattice interaction in the case of the hole localized at the
O-site and could suggest that the hole trapping is more preferential in the
oxygen sublattice. In both cases we deal with the several eV-effect both for $electronic$ and $ionic$ contributions to relaxation energy. Moreover, such an estimation seems to be typical for different insulators.\cite{Shluger,Vikhnin}
It is worth noting that the electron-lattice interaction is believed to be one of the main sources of correlated particle hopping resulting in different probabilities  for two types of a charge transfer.

\subsubsection{Generalized Peierls-Hubbard model and "negative-U" effect}

 Transition metal oxides with  strong electron and lattice polarization effects need in a revisit of many conventional theoretical concepts and approaches. In particular, we
 should modify conventional Hubbard model as it is done, for instance,  in a "dynamic" Hubbard
 model by Hirsch \cite{Hirsch1} or a modified Peierls-Hubbard model
 \cite{Nasu} with a classical description of the anharmonic core/shell displacements.
Having in mind the application to insulating cuprates let address a square lattice Hubbard model with a half-filling and a
strong on-site coupling of valent hole with core/shell displacements, which is described by the following Hamiltonian
$$
\hat H = -\sum _{<i,j>,\sigma} t(ij){\hat c}^{\dag}_{i\sigma}{\hat
c}_{j\sigma} +U\sum _{i,\sigma \sigma ^{'}}n_{i\sigma}n_{i\sigma
^{'}}
$$
\begin{equation}
+ \sum _{i}v_{an}(q_{i},n_{i}) + \sum _{<i,j>}v_{int}(q_{i},q_{j}),
\label{HHH}
\end{equation}
where ${\hat c}^{\dag}_{i\sigma}$ (${\hat c}_{j\sigma}$) are
creation (annihilation) operators for valent hole;
$t(ij)=t(q_{i},q_{j})$ is the transfer integral between two
neighboring lattice sites which depends on the dimensionless core/shell
displacement coordinate; $U$ is the on-site repulsion energy; $v_{an}(q_{i},n_{i})$ is the configurational energy  that incorporates the coupling between valent holes and the site-localized anharmonic core/shell mode with
dimensionless displacement coordinate $q_{i}$;
\begin{equation}
v_{an}(q_{i},n_{i})=
a(n_{i})q_{i}^{2}-b(n_{i})q_{i}^{4}+c(n_{i})q_{i}^{6}, \label{an}
\end{equation}
where $a,b,c$ are the functions of the hole occupation number such
as
\begin{equation}
a(n_{i})= a_{0}+a_{1}n_{i}+a_{2}n_{i}^{2} ,
\end{equation}
It is clear that $v_{an}(q_{i},n_{i})$ includes the renormalization
both of the one-particle energy and the on-site hole-hole repulsion.
The last term in (\ref{H}) represents the intersite configurational
coupling. The $q$-dependence of transfer integral implies the
correlated character of the hole hopping, and can be transformed
into the effective dependence on hole occupation number
\cite{Hirsch1}
\begin{equation}
t(n_{i},n_{j})= t(1+\alpha (n_{i}+n_{j})+\beta n_{i}n_{j}) .
\end{equation}
with $\alpha ,\beta$ being the correlated hopping parameters.

The conventional Hubbard Hamiltonian, or $t$-$U$-model, stabilizes
the spin density wave (SDW) electron order with $n_{i}= 1$. In a
strongly correlated limit $U\geq t$ the Hubbard model reduces to a
Heisenberg antiferromagnetic model. Depending on the parameters of
the hole-configurational coupling and correlated hopping the modified
Hubbard Hamiltonian (\ref{H}) can stabilize the
``disproportionated'' or charge ordered (CO) electron phase with the
on-site filling numbers $n=0$, and $n=2$ thus leading to the
``negative-U'' effect. Even simple modified model turns out to be
very complicated and leads to a very rich physics.\cite{Hirsch1}
Depending on the values of parameters the system yields the SDW
phase with no core/shell displacements as a $true$ ground state with a
global minimum of free energy, and CO phase with shell displacements
as a $false$ ground state with a local minimum, or vice
versa.\cite{Nasu} Strong anharmonicity $v_{an}(q_{i},n_{i})$ makes
possible phase transitions between the phases the first order ones.

\subsubsection{Vibronic reduction of charge transfer integrals}

In general, charge, spin and vibronic modes  are
strongly coupled and so we have to do with the hybrid modes. For a weak
intermode coupling regime the charge transfer is accompanied  by the
induced local structural  fluctuations, that provides the vibronic
reduction of the charge transfer integral:
\begin{equation}
t_{12}=t_{12}^{(0)}\,\ K_{vib}\,\quad
K_{vib}=\left\langle \chi _{1}|\chi _{2}\right\rangle ^{2},
\end{equation}
where \thinspace $K_{vib}$ is a vibronic reduction factor, $\left\langle \chi _{1}|\chi
_{2}\right\rangle $ is an overlap integral for the local oscillatory states
with and without particle transferred. In an opposite regime of the strong intermode coupling one assumes that
different electronic parameters for the $e$- and $h$-centers are
distinguished significantly up to different type of the adiabatic potential
and appropriate JT mode. This regime favors the charge localization.

The vibronic reduction factor $K_{vib}$, or the Franck-Condon factor
\cite{Bersuker} may be written as follows
\begin{equation}
K_{vib}=N\ exp(-\gamma ),
\end{equation}
where $N$ and $\gamma $ in a complicated manner depend on the vibronic constants, the oxygen and 3d-metal atomic
masses. For a simplest one-dimensional single-mode case
\begin{equation}
K_{vib}=\frac{2\tau }{1+\tau ^{2}}\exp \left( -\frac{\left( \Delta Q\right)
^{2}}{l_{1}^{2}+l_{2}^{2}}\right) ,
\end{equation}
where $l_{1}$ and $l_{2}$ are the effective oscillatory lengths of the $1$-
and $2$-centers, respectively, $\tau =l_{1}/l_{2}$, $\Delta Q$ is  the
distance  separating the minima of the adiabatic potential for the $1$- and $2$-centers.

\subsubsection{Spin reduction of charge transfer integrals}

Overall the paper we neglect a spin degree of freedom which can crucially impact on the charge transport. 
The most part of 3d oxides are characterized by an antiferromagnetic spin background that implies a localization effect due to strong spin reduction of one-particle transfer integrals and the  probability amplitude for   a polar center transfer $ M^{\pm} + M^0 \rightarrow M^0 +M^{\pm}$ or the motion of the electron (hole) center in the matrix of
$M^0$-centers. From the other hand, in antiferromagnets there is no problems with another type of the CT which specifies
the  probability amplitude for a  local disproportionation, or the spin-singlet $eh$-pair creation: 
$M^0 +M^0 \rightarrow M^{\pm}+M^{\mp}$, and the inverse process of the  spin-singlet  $eh$-pair recombination:
$M^{\pm}+M^{\mp}\rightarrow M^0 +M^0$. In other words, the  spin subsystem can strongly affect the correlated character of the charge transfer leading to unconventional situations like that of spin-singlet eh-pairs  moving through the lattice freely without disturbing the antiferromagnetic spin background, in contrast to the single particle motion. So, it seems that the situation in antiferromagnetic 3d insulators may  differ substantially from that in usual semiconductors or in other bandlike insulators where, as a rule, the effective mass of the electron-hole pair is larger than that of an unbound electron and hole.

\section{S=1 pseudospin formalism for model mixed valence system}

\subsection{Pseudospin operators}
The problem of the multi-stability of solids looks rather trivial when one say
about the orbital and/or spin degrees of freedom. Usually in such a case we
start from  the lattice of coupled orbital and/or spin  momenta described by
the relevant (spin-)Hamiltonian that implies the variety of possible collective
orbital and/or spin orderings that compete with each other under different
external conditions. In other words, the multi-stability accompanies the basic
degeneracy inherent to a certain atom, ion, or center with a nonzero orbital
and/or spin momentum. Such an outlook is believed to  be easily extended to
systems with charge degree of freedom which can be represented to be a system
of either centers which possible charge states form a pseudo-multiplet. Below
we address a simple
 model of a mixed-valence system with three possible stable valent states of a cation-anionic
 cluster, hereafter $M$: $M^0, M^{\pm}$, forming the charge (isospin) triplet. Starting from $M^0$ state as a bare vacuum state, we may address the $M^{\pm}$ centers as a result of pseudo-spin $\Delta S_z=\pm 1$ deviation, or as a hole and electron, respectively.
 Below we intend to concentrate themselves on charge degree of freedom, and
 that is why we neglect the orbital, spin, and lattice degrees of freedom. It implies a renormalization of different parameters, mainly it concerns the charge transfer. 
 
Similarly to  the neutral-to-ionic electronic-structural transformation
  in organic charge-transfer crystals (see paper by T. Luty in Ref.\onlinecite{Toyozawa})
   the system of  charge  triplets
can  be described in frames  of the S=1 pseudo-spin formalism.
  To this end we associate three charge states of the $M$-center with different valence:
  $M^0,M^{\pm}$  with three components of  $S=1$ pseudo-spin (isospin)
triplet with  $M_S =0,+1,-1$, respectively.

  The $S=1$ spin algebra includes three independent irreducible tensors
${\hat V}^{k}_{q}$ of rank $k=0,1,2$ with one, three, and five components,
respectively, obeying the Wigner-Eckart theorem \cite{Varshalovich}
\begin{widetext}
\begin{equation}
\langle SM_{S}| {\hat V}^{k}_{q}| SM_{S}^{'}\rangle=(-1)^{S-M_{S}} \left(
\begin{array}{ccc}S&k&S
\\
-M_{S}&q&M_{S}^{'}\end{array}\right)\left \langle S\right\| {\hat
V}^{k}\left\|S\right\rangle. \label{matelem}
\end{equation}
\end{widetext}
Here we make use of standard symbols for the Wigner coefficients and reduced
matrix elements. In a more conventional Cartesian scheme a complete set of the
non-trivial pseudo-spin operators would include both ${\bf   S}$ and a number
of symmetrized bilinear forms $\{S_{i}S_{j}\}=(S_{i}S_{j}+S_{j}S_{i})$, or
spin-quadrupole operators, which are linearly coupled to $V^{1}_{q}$ and $V^
{2}_{q}$, respectively
$$
V^{1}_{q}=S_{q}; S_{0}=S_{z}, S_{\pm}=\mp \frac{1}{\sqrt{2}}(S_{x}\pm iS_{y} ):
$$
\begin{equation}
V^{2}_{0} \propto (3S_{z}^{2}-{\bf  S}^2), V^{2}_{\pm 1}\propto (S_z S_{\pm}+
S_{\pm}S_z), V^{2}_{\pm 2}\propto S_{\pm}^2 .
\end{equation}
These pseudo-spin operators are not to be confused with real physical
spin-operators; they act in an imaginary pseudo-space.

To describe different types of  pseudo-spin ordering in a mixed-valence system
we have to introduce eight order parameters: two $diagonal$ order parameters
$\langle S_{z}\rangle$ and $\langle S_{z}^2\rangle$, and six {\it off-diagonal}
order parameters $\langle V^{k}_{q}\rangle$ ($q\not=0$). Two former order
parameters can be termed as $valence$ and $ionicity$, respectively. The {\it
off-diagonal} order parameters describe different types of the valence mixing.
Indeed, operators $V^{k}_{q}$ ($q\not=0$) change the $z$-projection of
pseudo-spin and transform the $|SM_S\rangle$ state into  $|SM_{S}+q\rangle$
one. In other words, these can change $valence$ and $ionicity$. It should be
noted that for the $S=1$ pseudospin algebra there are two operators:
$V^{1}_{\pm 1}$ and $V^{2}_{\pm 1}$, that change the pseudo-spin projection by
$\pm 1$, with slightly different properties
\begin{equation}
\langle 0 |\hat S_{\pm} | \mp 1 \rangle = \langle \pm 1 |\hat S_{\pm} | 0
\rangle =\mp 1, \label{S1}
\end{equation}
but
\begin{equation}
\langle 0 |(S_z S_{\pm}+ S_{\pm}S_z)| \mp 1 \rangle = -\langle \pm 1 |(S_z
S_{\pm}+ S_{\pm}S_z)| 0 \rangle =+1. \label{S2}
\end{equation}

\subsection{Gell-Mann operators and generalized pseudospin Hamiltonian}
Three spin-linear (dipole) operators $\hat S_{1,2,3}$ and five independent
spin-quadrupole operators $\{{\hat S_{i}},{\hat S_{j}}\}-\frac{2}{3} {\hat
{\bf  S}}^{2}\delta _{ij}$ given $S=1$ form eight Gell-Mann operators being the
generators of the SU(3) group. Below we will make use of the appropriate
Gell-Mann $3\times 3$
 matrices $\Lambda^{(k)}$,  which
differ from the conventional $\lambda^{(k)}$ only by a
renumeration:\cite{electr} $\lambda^{(1)}=\Lambda^{(6)}$,
$\lambda^{(2)}=\Lambda^{(3)}$, $\lambda^{(3)}=\Lambda^{(8)}$,
$\lambda^{(4)}=\Lambda^{(5)}$, $\lambda^{(5)}=-\Lambda^{(2)}$,
$\lambda^{(6)}=\Lambda^{(4)}$, $\lambda^{(7)}=\Lambda^{(1)}$,
$\lambda^{(8)}=\Lambda^{(7)}$. First three matrices $\Lambda^{(1,2,3)}$
correspond to linear (dipole) spin operators:
$$
\Lambda^{(1)}=S_x ;\quad  \Lambda^{(2)}=S_y ;
    \quad \Lambda^{(3)}=S_z
$$
while other five matrices correspond to quadratic (quadrupole)  spin operators:
$$
   \Lambda^{(4)}=-\{S_zS_y\} ;\quad
 \Lambda^{(5)}=-\{S_xS_z\} ;\quad
 \Lambda^{(6)}=-\{S_xS_y\} ;
$$
$$
\Lambda^{(7)}=-\frac{1}{\sqrt{3}}(S_x^2+S_y^2-2S_z^2) ; \quad
\Lambda^{(8)}=S_y^2-S_x^2 ;
$$
$$
S_x^2+S_y^2+S_z^2=2\hat E
$$
with $\hat E$ being a unit $3\times 3$ matrix.

The generalized spin-1 model can be described by the Hamiltonian  bilinear
 on the SU(3)-generators $\Lambda^{(k)}$
\begin{equation}
\hat{H}=-\sum_{i,\eta}\sum_{k,m=1}^{8}J_{km}\hat{\Lambda}_i^{(k)}
\hat{\Lambda}_{i+\eta}^{(m)} \, .\label{su3}
 \end{equation}
  Here $i,\eta$ denote lattice
sites and nearest neighbors, respectively. This is a $S=1$ counterpart of the
$S=1/2$ model Heisenberg Hamiltonian with three generators of  the SU(2) group
or Pauli matrices included instead of eight Gell-Mann matrices.

\subsection{Generalized mean-field model}
In frames of a classical, or mean-field description of the $S=1$ quantum
pseudo-spin system we start from a coherent state approximation with trial
functions \cite{electr}
\begin{equation}
 \psi=\prod_{j\in lattice}c_i(j)\psi_i=\prod_{j\in
lattice}(a_i(j)+ib_i(j))\psi_i . \label{wf}
\end{equation}
Here $j$ labels a lattice site and the spin functions $\psi_i$ in a Cartesian
 basis are used: $\psi_z=|10>$ and $\psi_{x,y}\sim(|11>\pm|1-1>)/\sqrt 2$.
 The linear (dipole) pseudo-spin operator within $|x,y,z\rangle$ basis is represented
  by a simple matrix:
 $$
 <\psi_i|S_j|\psi_k> =-i\varepsilon_{ijk},
 $$
  and for the order parameters one  easily obtains:
\begin{equation}
 <\hat{\bf  S}> = -2[{\bf  a} \times {\bf  b}]; \,
<\{\hat{S_i}\hat{S_j}\}>=2(\delta_{ij}-a_ia_j-b_ib_j) \label{med}
\end{equation}
given the normalization constraint ${\bf  a}^2 +{\bf  b}^2=1$. Thus, for the
case of spin-1 system the order parameters are determined by two classical
vectors (two real components of one complex vector $\bf  c =\bf  a +i\bf  b$
from (\ref{wf})). The two vectors are coupled, so the minimal number of dynamic
variables describing the $S=1$ spin system appears to be equal to four. Along
with $\bf  a $, $\bf  b$ vectors one might introduce ${\bf  l}=[\bf  a \times
\bf  b]$. Hereafter we would like to emphasize the $director$ nature of the
${\bf  c}$ vector field: $\psi ({\bf  c})$ and $\psi (-{\bf  c})$ describe the
physically identical states. It is worth noting that the coherent states
provide the optimal way both to a correct mean-field approximation (MFA) and
respective continuous models.\cite{electr}

\begin{figure}[ht]
\includegraphics[width=8.5cm,angle=0]{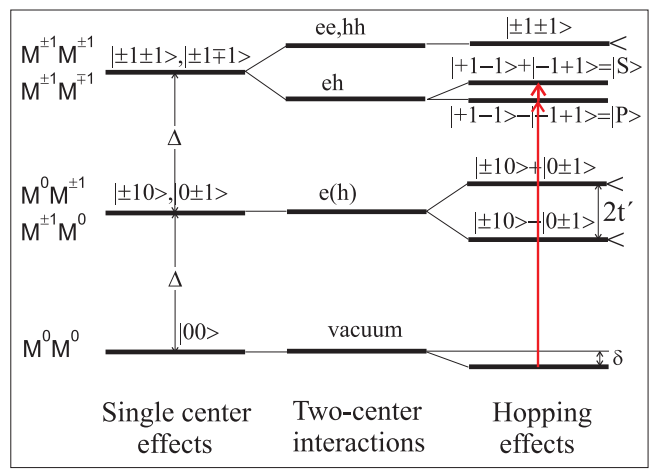}
\caption{Schematic energy spectrum of two $nn$ M-centers system (see text for details). Arrows mark the dipole-allowed CT transitions. }
\label{fig2}
\end{figure}

\section{Model mixed-valence system}

\subsection{Effective pseudo-spin Hamiltonian}
Effective pseudo-spin Hamiltonian for our model mixed-valence system should
incorporate a large number of contributions that describe different long- and
short-range coupling between $M^{0,\pm}$ centers, single-ion and two-ion terms.
Single-site terms can be subdivided into  {\it single-ion } anisotropy and {\it
pseudo-Zeeman} interaction. Bilinear and biquadratic two-site terms can be
subdivided into $diagonal$ interactions like "density-density", and {\it
off-diagonal} terms that describe charge fluctuations conserving the total charge of
the system, such as one-electron(hole) and two-electron(hole) transport. 
   An effective pseudo-spin
Hamiltonian of the model mixed-valence system which takes into
 account  the main part of aforementioned contributions can be represented as follows
$$
  \hat H =  \sum_{i}  (\Delta _{i}S_{iz}^2
  - h_{i}S_{iz}) + \sum_{<i,j>} v_{ij}S_{iz}^2 S_{jz}^2 +
  \sum_{<i,j>} V_{ij}S_{iz}S_{jz}+
$$
$$
\sum_{<i,j>} [D_{ij}^{(1)}(S_{i+}S_{j-}+S_{i-}S_{j+})+ 
D_{ij}^{(2)}(T_{i+}T_{j-}+T_{i-}T_{j+})]
$$
\begin{equation}
  +\sum_{<i,j>} t_{ij}(S_{i+}^{2}S_{j-}^{2}+S_{i-}^{2}S_{j+}^{2}),
  \label{H}
  \end{equation}
 where
 $$
 T_{\pm}= (S_z S_{\pm}+ S_{\pm}S_z).
 $$
Two first single-ion terms describe the effects of bare pseudo-spin splitting,
or the local energy of $M^{0,\pm}$ centers. Interestingly, the parameter $\Delta$ can be related with correlation Hubbard parameter $U$: $U=2\Delta$. The second term   may be
associated with an external, generally speaking, non-uniform pseudo-magnetic
field $h_i$, in particular, a real electric field. It is easy to see that it describes an electron/hole assymetry. The third and fourth terms describe the effects of long- and short-range inter-ionic interaction including screened Coulomb and covalent coupling. 

If to apply the familiar spin terminology, the first term in (\ref{H})
represents a single-ion anisotropy, the second does the Zeeman term, the fourth
and fifth do the anisotropic Heisenberg exchange, and the third and sixth do
the biquadratic spin-quadrupolar coupling.
To illustrate the role of different terms in (\ref{H}) we present in Fig.1 a schematic energy spectrum of $nn$ pair of $M$ centers provided an eh-symmetry ($h=0$) and $|00\rangle$ ground state ($\Delta >0$). It is worth noting the effect of a renormalization of  the ground state due to eh-pair creation/recombination effect ($t_{ij}^{\prime\prime}\not=0$) with a stabilization energy $\delta \approx |t_{ij}^{\prime\prime}|^2/2\Delta$. Two electron-hole states with S- (even) and P- (odd) type symmetry have a very strong
dipole coupling with the large value of $S-P$ transition dipole matrix element:
\begin{equation}
d = |\langle S|\hat{\bf d}|P\rangle | \approx 2eR_{MM} \approx 2e\times 4
\mbox{\AA}. \label{d}
 \end{equation}
Contrary to $P$-type pair state the $S$-type one  is dipole-forbidden and corresponds to
a so-called two-photon state. 

One should note that despite many simplifications, and first the neglect of
orbital and spin degrees of freedom,  quenched lattice approximation, the
effective Hamiltonian (\ref{H}) is rather complex, and represents one of the
most general forms of the anisotropic $S=1$ non-Heisenberg Hamiltonians. For
the system there are two classical ($diagonal$) order parameters: $\langle S_z
\rangle = n$ being a valence, or charge density with electro-neutrality
constraint $\sum _{i} n_i = \sum _{i} S_{iz}=0$, and $\langle S_{z}^{2} \rangle
= n_p$ being the density of polar centers $M^{\pm}$, or "ionicity". In
addition, there are two unconventional {\it off-diagonal} order parameters:
``fermionic'' $\langle S_+ \rangle $ and ``bosonic ''$\langle S_{+}^{2} \rangle
$; the former describes a phase ordering for the disproportionation reaction,
or the single-particle transfer, while the latter does for exchange reaction,
or for the two-particle transfer. Indeed, the ${\hat S}_+$ operator creates a
hole and is fermionic in nature, whereas the ${\hat S}_{+}^{2}$ does a hole
pair, and is bosonic in nature.

\subsection{Single and two-particle transport}
The last three terms in (\ref{H}) representing the one- and two-particle hopping, respectively, are of primary importance for the transport properties, and deserve special interest. 

Two types of one-particle hopping are
governed by two transfer integrals $D^{(1,2)}$, respectively.
 In accordance with (\ref{S1}) and
 (\ref{S2}) the transfer integral  $t_{ij}^{\prime}=(D_{ij}^{(1)}+ D_{ij}^{(2)})$
 specifies the  probability amplitude for a {\it local disproportionation, or the $eh$-pair creation}
$$
M^0 +M^0 \rightarrow M^{\pm}+M^{\mp};
$$
and the inverse process of the  {\it $eh$-pair recombination}
$$
M^{\pm}+M^{\mp}\rightarrow M^0 +M^0 ,
$$
while the transfer integral  $t_{ij}^{\prime\prime}=(D_{ij}^{(1)}- D_{ij}^{(2)})$
 specifies the  probability amplitude for   a polar center transfer
 $$
 M^{\pm} + M^0 \rightarrow M^0 +M^{\pm},
 $$
 or the {\it  motion of the electron (hole) center in the matrix of
$M^0$-centers} or motion of the $M^0$-center in the matrix of $M^{\pm}$-centers.
It should be noted that, if $t_{ij}^{\prime\prime}=0$ but $t_{ij}^{\prime}\not=0$, the eh-pair is locked in two-site configuration.

The two-electron(hole) hopping is governed by transfer integral
  $t_{ij}$, or  a probability amplitude for the exchange reaction:
 $$
M^{\pm}+M^{\mp}\rightarrow M^{\mp} +M^{\pm}.
$$
 or the {\it   motion of the electron (hole) center in the
matrix of hole (electron) centers}.

It is worth noting that in Hubbard-like models
 all the types of one-electron(hole) transport are governed by the same
 transfer integral: $t_{ij}^{\prime}=t_{ij}^{\prime\prime}=t_{ij}$, while our model implies independent parameters for a
 disproportionation/recombination process and simple quasiparticle motion
  in the matrix of $M^0$-centers. In other words, we deal with a "correlated"
  single particle transport.\cite{Hirsch} 

\subsection{Mean-field approximation:three generic MFA-phases} 
First of all we would like to emphasize
the difference between classical and quantum mixed-valence systems. Classical (or chemical)
description implies the neglect of the {\it off-diagonal} purely quantum CT
effects: $D^{(1,2)}=t=0$, hence the valence of any site remains to be definite:
$0, \pm 1$, and we deal with a system of localized polar centers. In quantum systems with a nonzero charge transfer we arrive at {\it
quantum superpositions of different valence states} resulting in $indefinite$
on-site valence and ionicity which effective, or mean values $\langle
S_{z}\rangle$ and $\langle S_{z}^2\rangle$ can vary from $-1$ to $+1$ and $0$
to $+1$, respectively.

Making projection of the effective pseudo-spin Hamiltonian for the system onto
a space of states like (\ref{wf}), we obtain an energy functional
 which equivalent to a classical energy of the two coupled vector  $({\bf  a
},{\bf  b})$ fields defined on the common lattice. Thus, in the  framework of
the pseudo-spin $S =1$ centers model when the collective wave
 function is represented to be a product of the site functions
 like (\ref{wf}), the quantum problem is reduced to a classical
variation problem for a minimum of the energy for two coupled vector fields.

In frames of mean-field approximation (MFA) we may make use of coherent states
(\ref{wf}) that provide a physically clear assignment of different phases with
a straightforward recipe of its qualitative and quantitative analysis. All the
MFA-phases one may subdivide into those with a definite and indefinite
ionicity, respectively. There are two MFA-phases with definite ionicity;

1) {\bf Insulating monovalent $M^0$-phase with $ \langle S_{z}^2\rangle =
0$:}

The $M^0$-phase is specified by a simple uniform arrangement of ${\bf  a}$ and ${\bf 
b}$ vectors parallel to $z$-axis: ${\bf  a}\parallel {\bf  b}\parallel O_z$. In
such a case the on-site wave function is specified by unit vector (${\bf  a}$,
or ${\bf  b}$) parallel to $z$-axis. It is a rather conventional ground state phase for
various charge transfer insulators such as oxides with a positive magnitude of $\Delta$ parameter ($U>0$). All the centers have the same bare $M^0$ valence state. In other words, the $M^0$-phase
is characterized both by definite site ionicity and valence.
 So,   all the order parameters turn into zero:
 $\langle S_{z}\rangle = \langle S_{z}^2\rangle =\langle S_{+}\rangle =
 \langle S_{+}^2\rangle =0$. This is an ``easy-plane'' phase for
the  pseudo-spins, but an ``easy-axis'' one for the  ${\bf  a}$ and/or ${\bf 
b}$ vectors.

2) {\bf Mixed-valence binary (disproportionated) $M^{\pm}$-phase with $ \langle
S_{z}^2\rangle = 1$:}

 This phase usually implies an overall disproportionation $M^0 +M^0 \rightarrow M^{\pm}+M^{\mp}$ that seems to be realizable if $\Delta$ parameter becomes negative one (negative $U<0$ effect). It is a rather unconventional
 phase for  insulators. All the centers have the  "ionized"  valence state, one half the $M^+$ state, and another half the $M^-$  one, though one may in common conceive of deviation from fifty-fifty distribution. A simplified "chemical" approach to $M^{\pm}$-phase as to a classical disproportionated
phase is widely spread in solid state chemistry.\cite{Ionov} 
 In contrast with the $M^0$ phase the $M^{\pm}$-phase is specified by a planar
orientation of ${\bf  a}$ and ${\bf  b}$ vectors (${\bf  a}, {\bf  b}\perp
O_z$) with a varied angle in between.
   There is no fermionic transport: $\langle S_{+}\rangle =
0$,  while the bosonic one may exist, and, in common, $\langle S_{+}^2\rangle
=-\cos(\phi _{a}-\phi _{b})e^{i(\phi _{a}+\phi _{b}) }\not= 0$. This is an
``easy-axis'' phase for the pseudo-spins, but an ``easy-plane'' one for the
${\bf  a}$ and ${\bf  b}$ vectors.

The mixed valence $M^{\pm}$ phase as a system of strongly correlated electron  and hole  centers is equivalent to the lattice hard-core Bose system with an inter-site repulsion, or {\it electron-hole Bose liquid} (EHBL) in contrast with EH liquid in conventional semiconductors like Ge, Si where we deal with a two-component {\it Fermi-liquid}. Indeed, one may  address the electron $M^-$ center  to be a system of a local  boson ($e^2$) localized on the hole  $M^+$ center: $M^- = M^+ + e^2$.

In accordance with this analogy we assign three well known  molecular-field uniform phase
states of the $M^{\pm}$ binary mixture:

i) {\bf  charge ordered (CO) insulating state} with $\langle S_{z}\rangle = \pm
1$, ${\bf  a}\perp {\bf  b}$, and zero modulus of bosonic off-diagonal order parameter:
$|\langle S_{+}^2\rangle |= 0$;

ii) {\bf Bose-superfluid (BS) superconducting state} with $\langle S_{z}
\rangle
 = 0$,  ${\bf  a}$ and $ {\bf  b}$ being collinear, $\langle S_{+}^2\rangle = e^{2i\phi}$;

iii) {\bf mixed Bose-superfluid-charge ordering  (BS+CO) superconducting state
(supersolid) }
 with $0<|\langle S_{z}\rangle |<1$,  ${\bf  a}$ and ${\bf  b}$ being
oriented in $xy$-plane, but not collinear, $\langle S_{+}^2\rangle =
 -\cos(\phi _{a}-\phi _{b})e^{i(\phi _{a}+\phi _{b})}\not= 0$.

 In addition, we should mention the high-temperature non-ordered (NO) Bose-metallic phase
 with $\ll S_z \gg =0$.
 
 Rich phase diagram of $M^{\pm}$ binary mixture with unconventional superfluid and supersolid regions looks tempting, however, actually, their stabilization requires strong suppression of Coulomb repulsion between electron (hole) centers. Despite significant screening effect, the stabilization of uniform BS or BS+CO superconducting state as a result of a disproportionation reaction in a bare insulator \cite{Ionov} seems to be unrealistic.

3) {\bf Mixed-valence ternary (``under-disproportionated'') $M^{0,\pm}$-phase:}

For two preceding cases the order parameter $\langle S_{z}^2\rangle $, or
ionicity relates to its limiting values (0 or 1, respectively). For the
MFA-phase with indefinite ionicity, or  mixed-valence ternary
(``under-disproportionated'') $M^{0,\pm}$-phase,  $0 < \langle S_{z}^2\rangle
<1$, that is we have  a mixture of the $M^0,M^{\pm}$ centers. From the
viewpoint of the classical ${\bf  a}, {\bf  b}$ vectors formalism the phase
corresponds to the arbitrarily space-oriented ${\bf  l}=[{\bf  a}\times {\bf 
b}]$ vector. Both off-diagonal order parameters, fermionic $\langle
S_{+}\rangle $ and bosonic $\langle S_{+}^2\rangle $ are, in common, non-zero,
albeit with some correlation in between. So, for the ternary system one expects
a coexistence of fermionic and bosonic  transport.

It should be noted that a complete pseudo-spin description of the two last
model mixed-valence systems implies a two-sublattice approximation to be a
minimal model compatible with a sign of Coulomb interaction and a respective
tendency towards the checkerboard-like charge ordering.

\section{Insulating monovalent $M^0$-phase}
Insulating monovalent $M^0$-phase is a typical one for the ground state of insulating transition metal oxides, or Mott-Hubbard insulators. It is worth noting that in frame of conventional band model approach the $M^0$-phase, e.g., in parent cuprates, is associated with a metallic half-filled hole band system.  Below we address different types of quasiparticle excitations in such a system focusing on the features of the correlated hopping, governed generally by two transfer integrals $t_{ij}^{\prime}$ and $t_{ij}^{\prime\prime}$ which competition results in unconventional properties of electrons and holes in bare insulating monovalent $M^0$-phase. We show that $M^0$-phase can be unstable with regard to the self-trapping of CT excitons and nucleation of droplets of EH Bose liquid.  

\subsection{Electron, hole, and electron-hole excitations}

Starting from monovalent $M^0$-phase as a vacuum state $\left|0\right\rangle$ we introduce an electron-hole representation where $M^-_i$ center is derived as a result of an electron creation $\hat a_i^\dagger \left|0\right\rangle$, and $M^+_i$ center is derived as a result of a hole creation $\hat b_i^\dagger \left|0\right\rangle$. Then we transform pseudo-spin Hamiltonian (\ref{H}) into that of a system of effective electrons and holes
$$
\hat H= \Delta \sum _{i}(n^h_i+n^e_i)+\sum_{\left\langle ij\right\rangle}t_{ij}(\hat a_i^\dagger \hat a_j+\hat b_i^\dagger \hat b_j)+
$$
$$
	 \sum_{i,j}[V^{hh}(ij)(n^h_in^h_j+V^{ee}(ij)n^e_in^e_j+V^{eh}(ij)(n^h_in^e_j+n^h_jn^e_i)]
$$
\begin{equation}
+\sum_{\left\langle ij\right\rangle}t_{ij}^{\prime \prime}(\hat a_i^\dagger \hat b_j^\dagger+\hat a_i \hat b_j)
\label{eh}
\end{equation}
where $V_{ii}\rightarrow\infty$ to prohibit two-particle occupation of a single site. Here we suppose $h=0$ that provides  an electron-hole symmetry. The first line in (\ref{eh}) represents the single particle (electrons/holes) terms, the second one does the interparticle coupling, while the third one describes the creation and annihilation (recombination) of eh-pairs. It is worth noting that the latter terms describe some sort of eh-coupling. 

In terms of a pseudo-spin analogy the electrons and holes are associated with pseudo-spin $\Delta S_z=\pm 1$ deviations for an easy-plane magnet, localized or delocalized (pseudo-spin wave).

The behavior of electron/hole system crucially depends on the relation between two transfer integrals $t_{ij}^{\prime},t_{ij}^{\prime \prime}$
Below we address two distinct limiting  situations:

I. $t_{ij}^{\prime \prime}=0$: Forbidden recombination/creation  regime. In this case we deal with the bands of well defined electrons and holes with a charge gap $E_g^{e,h}=\Delta - z|t_{nn}^{\prime}|$ for both types of carriers.  Optical gap for unbound eh-pairs is $E_g^{eh}=2E_g^{e,h}$. However, in such a case we may expect for the formation of
Wannier excitons, or  eh-pairs bound due to a screened Coulomb eh-coupling. In terms of pseudo-spin formalism  the Wannier excitons may be regarded as two pseudo-spin waves having formed a quasilocalized state due to a long-range antiferromagnetic Ising exchange: $V_{ij}S_{iz}S_{jz}$.

II. $t_{ij}^{\prime}=0$: Regime of localized  electrons and holes with a dimerization effect and well defined $nn$ eh-pairs, or CT excitons. In such a case the charge gap is $E_g^{e,h}=\Delta $ for both types of localized quasiparticles. As it is clearly seen in Fig.1 the $t_{ij}^{\prime \prime}\not=0$ hopping results in a dimerization effect with a quantum renormalization of the vacuum state and indefinite ionicity, the formation of two types of localized eh-pairs, or CT excitons of S- and P-type. In frames of our $nn$ approximation the CT excitons are localized.
Optical gap is determined by the energy of P-type CT exciton: $E_g^{eh}=2\Delta+\delta$. 

Thus we arrive at two limiting types of monovalent $M^0$ insulators with a dramatic difference in behavior of electrons and holes, as well as electron-hole pairs. In type I insulators ($t_{ij}^{\prime}\gg t_{ij}^{\prime \prime}$) we deal with well defined bands of electrons and holes forming Wannier excitons, while in type II insulators ($t_{ij}^{\prime\prime}\gg t_{ij}^{\prime}$) we deal with localized electrons and holes which can form $nn$ eh-pairs, or CT excitons. 

The most part of 3d oxides are characterized by an antiferromagnetic spin background that implies strong spin reduction of one-particle transfer integrals $t_{ij}^{\prime}$. In other words, these, seemingly, belong to type II insulators, where  spin-singlet CT excitons can move through the lattice freely without disturbing the antiferromagnetic spin background, in contrast to the single hole motion. So, it seems that the situation in antiferromagnetic 3d insulators differs substantially from that in usual semiconductors or in other bandlike insulators where, as a rule, the
effective mass of the electron-hole pair is larger than that of an unbound electron and hole.

The Wannier excitons are formed due to an eh-coupling, while the CT excitons are formed due to a {\it kinetic cutoff}, or a specific feature of correlated hopping, in other words, the former have a $potential$, while the latter a $kinetic$ nature. 

It is worth noting that both M-centers within P-type CT excitons have a certain ionicity in contrast to S-type CT excitons which can mix with bare $M^0M^0$ ground state. CT excitons form peculiar quanta of a disproportionation reaction and may be viewed to be a minimal droplet of electron-hole Bose liquid.

In general, eh-excitations in $M^0$ phase consist of superpositions of pairs of free electrons and holes, and CT excitons. One expects two types of superpositions: CT exciton-like and band-like. The former have a localized character, while the latter have an itinerant one. 

The nature of the optical excitations accompanied by creation of electron-hole
 pairs  in 3d oxides is not fully understood.
 One of the central issues in the analysis of electron-hole
excitations  is whether low-lying states   are comprised of free
charge carriers or excitons. A conventional approach implies that if
the Coulomb interaction is effectively screened and weak, then the
electrons and holes are only weakly bound and move essentially
independently as free charge-carriers. However, if the Coulomb
interaction between electron and hole is strong, excitons are believed to
form, i.e.\ bound particle-hole pairs with strong correlation of
their mutual motion.  
 
 One of the most popular criteria to discriminate between the states relates to the band
 gap: states below the charge gap correspond to excitons with binding energy
 $ E_b = E_g - E$, and states above the charge
 gap do to free electron-hole pairs. However, this criteria seems to be oversimplified, and the states should be characterized as bound or unbound according to the
 scaling of the average electron-hole separation with system size. Excitons are
 entities with small electron-hole separation which remain finite as the system
 size is increased. By contrast, the average separation between two free charge
 carriers increases indefinitely with system size.
 To distinguish bound and unbound electron-hole states one might use the
 density-density correlation function \cite{Boman}
 $$
 C(i,j)=\langle ({\hat n}_{i}-\langle {\hat n}_{i}\rangle )
 ({\hat n}_{j}-\langle {\hat n}_{j}\rangle )\rangle ,
 $$
which measures a correlation of charge fluctuations on site $i$  to a
charge fluctuations on site $j$. A negative value correlates an excess
(deficit)  of charge with deficit (excess), or electron-hole distribution. In frame of pseudo-spin approach this 
correlation function measures the longitudinal ($\parallel z$) short-range antiferromagnetic fluctuations 
$$
 C(i,j)=\langle \hat S_{iz}\hat S_{jz}\rangle 
 -\langle \hat S_{iz}\rangle \langle \hat S_{jz}\rangle .
 $$
  The results of the direct
 computations for conjugated polymers in frames of the extended Hubbard model
 \cite{Boman} show for different symmetry sectors that there exist unbound
 states below the charge gap, and bound states above  the charge gap. Thus,
 the charge gap, often used to define the binding energy of excitons, is not a
  decisive  criterion by which to decide whether a state is bound or unbound.

In practice, many authors consider excitons to consist of real-space
configurations with electrons and holes occupying the nearest neighbor sites,
while the electrons and holes are separated from each other in the
conduction-band states \cite{Guo}.

  The relative energy position and transition intensity of excitons and free unbound
  electron-hole pairs is an issue of large complexity. For instance,
  in polydiacetylenes \cite{Guo} the large absorption peak at $1.85$ eV is attributed
  to an exciton, since photoconductivity is absent in this energy region. The latter
  has a threshold at around $2.4$ eV. Interestingly, that observed photoconductivity band
  is not visible in the conventional linear (one-photon) absorption spectrum.
  Because of the strong oscillator strength of the exciton, the conduction band has a
  weak  oscillator strength and is enveloped by the high-energy tail of the exciton peak.

\subsection{Coupling with electromagnetic field}

The electron-hole excitations and optical properties of strongly correlated
  electron systems is of current  both experimental and theoretical  interest. In particular, the optical conductivity is  of fundamental theoretical interest because the spectral weight at low frequencies seems to be the natural order parameter for the Mott transition. 
The great body  of experimental information regarding the  eh-excitations in solids is provided by optical and electron energy loss (EELS) spectroscopy, however, its interpretation depends crucially on the theoretical scheme used.
The optical conductivity of different model systems has been studied
by approximate mean field calculations, by analysis of integrable  1D models,
 by exact diagonalization of small
 systems, and by quantum Monte Carlo techniques. The uncertain quantitative
 applicability of the analytic mean field calculations,
 the size limitations of the exact diagonalization and Monte Carlo results
and the possibility that the integrable 1D models do not exhibit generic
behavior one lead  to consider other methods for obtaining information about
the optical conductivity.
An outgoing beyond the effective Hamiltonian methods with restricted basis
which are relevant for  description of the lowest energy excitations in the
extremely limited range of energies, and the elaboration of effective methods
to describe optically excited states is a challenging problem in solid state
physics.

 Making use of a standard Peierls transformation in hopping terms in (\ref{eh}) we arrive at an
effective Hamiltonian for the coupling with electromagnetic field   
\begin{eqnarray}
{\hat t_{ij}} \rightarrow {\hat t_{ij}}e^{i(\Phi _{j}-\Phi _{i})},
\end{eqnarray}
\begin{eqnarray}
(\Phi _{j}-\Phi _{i})=-\frac{q}{\hbar c}\int _{{\bf  R}_{i}}^{{\bf  R}_{j}}{\bf  A}({\bf  r})d{\bf  r},
\end{eqnarray}
where  ${\bf  A}$ is the vector-potential, and integration runs over line binding the  $i$ and $j$ sites. For the nearest neighbours one may use the simplified relation
\begin{eqnarray}
(\Phi _{j}-\Phi _{i})=-\frac{q}{\hbar c}({\bf  A}({\bf  R}_{i})\cdot {\bf  R}_{ji}).
\end{eqnarray}
 
The  current density operator one may represent to be a sum of three terms
\begin{eqnarray}
{\bf j}^{ee}({\bf R}_{i})=\frac{iq}{2\hbar }\sum _{j}t_{ij}{\bf R}_{ij}(\hat a_i^\dagger \hat a_j-\hat a_i\hat a_j^\dagger ),\label{1}
\end{eqnarray}
\begin{eqnarray}
{\bf j}^{hh}({\bf R}_{i})=\frac{iq}{2\hbar }\sum _{j}t_{ij}{\bf R}_{ij}(\hat b_i^\dagger \hat b_j-\hat b_i\hat b_j^\dagger ),\label{2}
\end{eqnarray}
\begin{eqnarray}
{\bf j}^{eh}({\bf R}_{i})=\frac{iq}{2\hbar }\sum _{j}t_{ij}{\bf R}_{ij}(\hat a_i^\dagger \hat b_j^\dagger -\hat b_i\hat a_j ),\label{3}
\end{eqnarray}
where the two first terms describe the electron and hole currents, respectively, while the third term describe the current fluctuations due to eh-pair creation/annihilation. 

Standard linear-response theory then yields for optical
conductivity ($T=0$)
\begin{equation}
{\mbox Re}\sigma ({\bf  q},\omega)=(\pi /\omega) \sum _{e} |\mu _{ge}({\bf  q})|^2
\delta (E_{e}-E_{g}-\hbar \omega),
\end{equation}
where
$$
\mu _{ge}({\bf  q}) = \langle \Psi _{g}|j_{{\bf  q}}|\Psi _{e}\rangle
$$
is a transition matrix element, and sum runs over all the excited $\Psi _{e}$ states. 

The first two current density operators (\ref{1})-(\ref{2}) describe electron and hole intraband transition so that
 the optical absorption spectra in $M^0$ phase are specified only by  the latter term (\ref{3}). In other words, the optical response in $M^0$ phase of our model system is determined only through a CT exciton channel. 
 
 It should be noted that the electron/hole current operators in (\ref{1})-(\ref{2}) can be expressed through a superposition of  antisymmetric pseudospin operators:
\begin{equation}
(\hat S_{i+}\hat S_{j-}-\hat S_{i-}\hat S_{j+})-\hat T_{i+}\hat T_{j-}-\hat T_{i-}\hat T_{j+}),
\end{equation}
while eh-current fluctuation operator in (\ref{3}) can be expressed through another superposition of the same antisymmetric pseudospin operators
\begin{equation}
(\hat S_{i+}\hat S_{j-}-\hat S_{i-}\hat S_{j+})+\hat T_{i+}\hat T_{j-}-\hat T_{i-}\hat T_{j+}).
\end{equation}

\subsection{Charge transfer instability and CT exciton self-trapping}
Electron and hole in a CT exciton in type II $M^0$ insulator  are strongly coupled both in between and with the lattice.
In contrast with conventional wide-band semiconductors where the excitons
dissociate easily producing two-component electron-hole gas or plasma,
\cite{Rice}  small CT excitons  both free and self-trapped  are likely
   to be stable with regard the eh-dissociation. 
To illustrate the principal features of CT exciton self-trapping effect we address a simplified two-level model of a two-center MM cluster in which a ground state and a CT state are associated with a pseudospin 1/2 doublet, $|\uparrow\rangle$ and $|\downarrow\rangle$, respectively. In addition,  we introduce some configurational coordinate $Q$ associated with a deformation of the cluster, or respective anionic background.\cite{shift} Such a model is typical one for so-called (pseudo)Jahn-Teller systems.
As a  starting point of the model we introduce the effective electron-configurational Hamiltonian  as follows
\begin{equation}
H_{s}= -\Delta {\hat s}_{z} -t{\hat s}_{x}- pQ +\frac{K}{2}Q^2 - aQ{\hat s}_{z}\, , \label{HH}	
\end{equation}
where $s_z = \frac{1}{2}$\begin{math}\left(
\begin{array}{clrr} % 
1 &  0  \\ 
0 & -1   
 \end{array}\right)
 \end{math}, 
 $s_x = \frac{1}{2}$\begin{math}\left(
\begin{array}{clrr} % 
0 & 1  \\ 
1 & 0   
 \end{array}\right)
 \end{math}
 are  Pauli matrices, and the first term describes the bare energy splitting of "up" and
"down"  states with energy gap $\Delta $ (CT energy), while the second term describes the coupling (mixing) between "up" and "down" pseudospin states.
 In terms of a pseudospin analogy the both parameters may be associated with effective fields.  The
third and fourth terms in (\ref{H}) describe the linear and quadratic contributions to the configurational energy. Here, the linear term formally corresponds to the energy  of an external configurational strain  described by an effective strain parameter $p$, while quadratic term with "elastic" constant $K$ is associated with a conventional harmonic approximation for  configurational energy. The last term describes the electron-configurational (vibronic)
interaction with $a$ being a electron-configurational coupling constant. Hereafter we make use of dimensionless configurational variable $Q$ therefore all of the model parameters are assigned the energy units. 
Our model Hamiltonian has the most general form except the simplified form of electron-configurational coupling where we omit the term $\propto Q {\hat s}_x$.
In frame of adiabatic approximation the eigenvectors for the Hamiltonian can be written as follows:
$$
\Psi _+(Q) =\cos\alpha |\uparrow\rangle + \sin\alpha
|\downarrow\rangle ;
$$
 \beq
 \Psi _-(Q) =\sin\alpha |\uparrow\rangle -
\cos\alpha |\downarrow\rangle , \eeq where
$$
\tan2\alpha=\frac{t}{\Delta +aQ}\, .
$$
 The corresponding eigenvalues
\beq
E_{\pm}(Q)=\frac{K}{2}Q^2 - pQ \pm\frac{1}{2}\left[(\Delta +aQ)^2 + t^2
\right]^{1/2}
\eeq
define the upper and lower branches of the  configurational, or adiabatic potential (AP),
respectively. These potential curves describe the energy of $|\pm\rangle$ states as functions of configurational coordinate $Q$. 
\begin{figure}[t]
\includegraphics[width=8.5cm,angle=0]{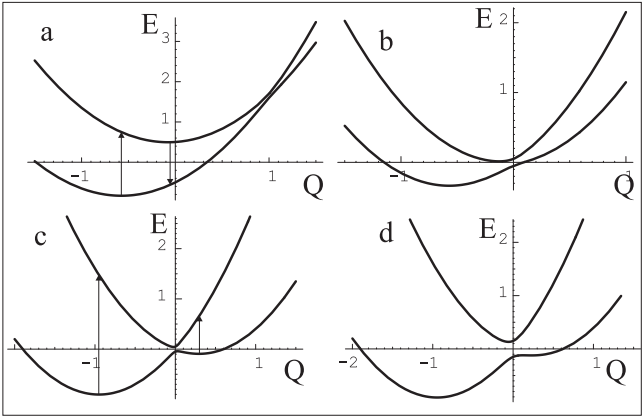}
\caption{Typical adiabatic potentials  of CT center: a) $\Delta =1.0; t=0.1; p=-0.65; a=1.0; K=2.0$; b)$\Delta =0.0; t=0.1; p=-0.65; a=1.0; K=2.0$; c)$\Delta =0.0; t=0.1; p=-0.65; a=2.5; K=2.0$; d)$\Delta =0.0; t=0.3; p=-0.65; a=2.5; K=2.0$. Arrows mark CT transitions (see text for details).} \label{fig3}
\end{figure}
The impact of different model parameters on the shape of AP can be easily understood, if we neglect transfer integral ($t =0$) and deal with two ideal parabolas describing the configurational energy for "up" and "down" states $|\uparrow ,\downarrow\rangle$, respectively. These identical parabolas with minima at
$Q_{\pm}^{(0)}=(p\mp \frac{a}{2})/K$ are shifted relative to each other. 
 The shift in
configurational coordinate is $\Delta Q = Q_{+}^{(0)} -Q_{-}^{(0)}
=a/2K$, while the shift in energy is $\Delta E^{(0)} = E_{+}^{(0)}
-E_{-}^{(0)} =pa/K+\Delta$, where
$E_{\pm}^{(0)}=E_{\pm}(Q_{\pm}^{(0)})$ or the energy of the bottoms
of respective parabolas. Interestingly, the energy shift is determined both by "mechanic" and "electronic" forces.
It is worth noting that the $Q$-shift of the "center of gravity" of AP is determined by the effective strain $p$. The electron-configurational coupling leads to a pulling apart the parabolas for "up" and "down" electronic states.  
The condition \beq |\Delta E^{(0)}| =
|pa/K+\Delta | = a^2/2K \label{LU}\eeq defines two specific points, where the minimum of one of the branches
crosses another branch, thus specifying the parameters range  admissive of a $bistability$ effect. If $E_{+}^0 >E_{-}(Q_{+}^{(0)})$ we arrive at rather conventional situation which is typical for long-lived CT states capable of decaying due to spontaneous Franck-Condon  radiative transitions. In opposite case $E_{+}^0 <E_{-}(Q_{+}^{(0)})$ we deal with  only  radiativeless relaxation channels, and the situation strongly depends on the magnitude of transfer  integral $t \not=0$ which leads to a crucial rearrangement of AP  near the crossection point of two branches with a number of novel effects of principal importance. First, we obtain two isolated branches of AP, the upper and lower ones, respectively,  with  quantum superpositions $\Psi _{\pm}$ which describe unconventional  states with a mixed valence of CuO$_4$ plaquettes. 
An illustrative example of typical adiabatic potentials is shown in Fig.3. Upper branch of AP has a single minimum point S with approximately "fifty-fifty" composition  $|\uparrow ,\downarrow\rangle$ of the respective superposition state $\Psi _{+}$.  

  For lower branch of AP we have either a single minimum point or  the two-well structure with two local minimum points (see Fig.3c), leading to a "bistability" effect which is of primary importance for our analysis. Indeed, these two points may be associated with two (meta)stable charge states with and without CT, respectively,  which form two candidates to struggle for a ground state.

 It is easy to see, that for large values of the transfer  integral  the system does not manifest bistability, i.e. the lower branch of conformational potential has only a single minimum (see, e.g., Figs.3c,d).
 
 Thus we  conclude that all the systems such as copper oxides may be divided to two classes: {\it CT stable systems} with the only lower AP branch  minimum for a certain  charge configuration, and bistable, or {\it CT unstable systems} with two lower AP branch minima for two local charge configurations one of which is associated with self-trapped CT excitons  resulting from self-consistent charge transfer and electron-lattice relaxation.\cite{Vikhnin}  
 Moreover, in the latter case we deal with an additional   metastable state on the upper AP branch which is characterized by a mixed valence and ionicity and seems to be intermediate one. 
 
 Our simple model analysis allows to uncover the relative role of different parameters governing the charge stability in the system. At the same time one should note that actually we deal with more complex multimode problem with a large body of intermediate metastable states which can be seen in luminescence spectra.

 \subsection{Nucleation of EH droplets and phase separation effects in CT unstable $M^0$ phase}
The AP bistability in CT unstable insulators points to tempting perspectives  of their evolution under either external 
impact.  
Metastable CT excitons in the CT unstable $M^0$ phase being the disproportionation quanta present candidate
 "relaxed excited states" to struggle for stability with ground state and the natural nucleation centers for electron-hole liquid phase. What way the CT unstable $M^0$ phase can be transformed into novel phase?  It seems likely that such a phase transition
could be realized due to a mechanism familiar to semiconductors with filled bands such as Ge and Si where
  given  certain conditions one observes a formation of metallic EH-liquid as a result of the exciton decay.\cite{Rice}
 However, the  system of strongly correlated electron $M^{-}$ and hole
$M^{+}$ centers  appears to be equivalent to an electron-hole
Bose-liquid  in contrast with the electron-hole  Fermi-liquid in
conventional semiconductors. However, the Wannier excitons in the latter wide-band systems
dissociate easily producing two-component electron-hole gas or plasma,\cite{Rice} while 
 small CT excitons  both free and self-trapped  are likely
   to be stable with regard the el-h-dissociation. At the same time, the two-center
    CT excitons have a very large fluctuating electrical dipole moment
     $|d|\sim 2eR_{MM}$ and can
be involved into attractive electrostatic dipole-dipole interaction. Namely
this is believed to be important incentive to the proliferation of excitons and
its clusterization. The CT excitons are proved to attract and form molecules
called biexcitons, and more complex clusters where the individuality of the
separate exciton is likely to be  lost. Moreover, one may assume that  like the semiconductors with indirect band gap structure, it is energetically favorable for the system to separate into a low density exciton
phase coexisting with the microregions of a high density two-component phase
composed of electron $M^-$ and hole $M^+$ centers, or EH droplets. 
Indeed, the excitons may be considered to be well defined entities only at
small content, whereas at large densities their coupling is screened and their
overlap becomes so considerable  that they loose individuality and we come to
the system of electron $M^-$ and hole $M^+$ centers, which form a  electron-hole Bose liquid.
An increase of
injected excitons in this case merely increases the size of the EH droplets,
without changing the free exciton density.

{\it Homogeneous nucleation} implies the spontaneous formation of EH droplets
due to the thermodynamic fluctuations in exciton gas. Generally speaking, such
a state with a nonzero volume fraction of EH droplets and the spontaneous
breaking of translational symmetry can be stable in nominally pure insulating
crystal. However, the level of intrinsic non-stoihiometry in 3d oxides is
significant (one charged defect every 100-1000 molecular units is common). The
charged defect produces random electric field, which can be very large (up to
$10^8$ Vcm$^{-1}$) thus promoting the condensation of CT excitons and the
{\it inhomogeneous nucleation} of EH droplets.
 Deviation from the neutrality  implies the
existence of additional electron, or hole centers that can be the natural
 centers for the {\it inhomogeneous nucleation} of the EH droplets.
Such droplets are believed to provide the more effective screening of the
electrostatic repulsion for additional electron/hole centers, than the parent
insulating phase. As a result, the electron/hole injection to the insulating
$M^0$ phase due to a nonisovalent substitution as in La$_{2-x}$Sr$_x$CuO$_{4}$,
Nd$_{2-x}$Ce$_x$CuO$_{4}$, or change in oxygen stoihiometry as in
YBa$_2$Cu$_3$O$_{6+x}$, La$_{2}$CuO$_{4-\delta}$,
La$_{2}$Cu$_{1-x}$Li$_x$O$_{4}$, or field-effect is believed to shift the phase
equilibrium from the insulating state to the unconventional electron-hole Bose
liquid, or in other words  induce the insulator-to-EHBL phase transition.

The optimal way to the nucleation of EH droplets  in parent system like
La$_2$CuO$_4$, YBa$_2$Cu$_3$O$_6$ is to create charge inhomogeneity by
nonisovalent chemical substitution in CuO$_2$ planes or in ``out-of-plane
stuff'', including interstitial atoms and vacancies. This process results in an
increase of the energy of the parent phase and creates proper conditions for
its competing with others phases capable to provide an effective screening of
the charge inhomogeneity potential. The strongly degenerate system of electron
and hole centers in EH droplet is one of the most preferable ones for this
purpose. At the beginning (nucleation regime) an EH droplet nucleates as a
nanoscopic cluster composed of several number of neighboring electron and hole
centers pinned  by disorder potential. Charged defects  supporting the EH droplet nucleation promote the formation of metastable ("superheated") clusters of parent phase.  It is clear that such a situation does not exclude the self-doping with the formation of a self-organized collective charge-inhomogeneous state in systems
which are near the charge instability.

EH droplets  can manifest itself remarkably  in various properties of the
3d oxides even  at  small volume fraction, or in a ``pseudo-impurity regime''. Insulators in a PS regime should be considered as phase inhomogeneous systems with, in general, thermo-activated mobility of the inter-phase
boundaries.  On the one hand, main features of this ``pseudo-impurity regime'' would be determined by the
partial intrinsic contributions of the appropriate phase components with
possible limitations imposed by the finite size effects. On the other hand, the
real properties will be determined  by the peculiar geometrical factors such as
a volume fraction, average size of droplets and its dispersion, the shape and
possible texture of  the droplets, the geometrical relaxation rates. These
factors are tightly coupled, especially near phase transitions for either phase
(long range antiferromagnetic ordering for the parent phase, the charge
ordering and other phase transformations  for the EH droplets) accompanied  by
the variation in a relative volume fraction.  Numerous
examples of the unconventional behavior of the 3d oxides in the pseudo-impurity
regime could be easily explained with taking into account the inter-phase
boundary effects (coercitivity, the mobility threshold, non-ohmic conductivity, oscillations,
relaxation etc.) and corresponding characteristic quantities.
Under increasing doping the ``pseudo-impurity regime'' with a relatively small
volume fraction of EH droplets (nanoscopic phase separation) can gradually transform into a  macro- (chemical) ``phase-separation regime'' with a sizeable volume fraction of EH droplets, and finally to a new EH liquid phase.  Phase separation is now widely discussed as an important phenomenon accompanying the high-T$_c$ superconductivity.

 Our scenario can readily explain photo-induced effects in pseudo-impurity phase.\cite{PIPS}  Indeed, the illumination of a material with light leads to the generation of eh-pairs that will proliferate and grow up to be a novel nonequilibrium EH droplet or simply to be trapped in either EH droplet with the rise in its volume fraction. The excitation energy appears to be lower when exciton is closer to the EH droplet. Therefore once the excitation transfer is finite, the optical excitation is attracted to the nearest neighbour of the EH clusters so that this cluster expands effectively under the light irradiation. In other words, the photoexcitation would result in an increase of the EH droplet volume fraction, that is why its effect in optical response resembles that of chemical doping. After switching off the light the droplet phase would relax to the thermodynamically stable state. Furthermore, such a simple  model can immediately explain the persistent photoconductivity (PPC) phenomena, found in insulating YBaCuO system,\cite{Kirilyuk} where the oxygen reodering provides the mechanism of a long-time stability for the EH droplets. In PPC phenomena, an illumination of a material with light leads to a long-lived photoconductive state. During the illumination of underdoped YBaCuO near the insulator-metal transition, the  material may even become superconducting.

\section{Electron-hole Bose liquid}
 Hereafter we would like to address the bosonic nature and some properties of the mixed-valence
 binary (disproportionated) $M^{\pm}$-phase termed as electron-hole Bose liquid (EHBL) in more details.
 For such a system, the pseudo-spin Hamiltonian (\ref{H}) can be mapped
 onto the Hamiltonian of hard-core Bose gas on a lattice (Bose-Hubbard model) which standard form
  can be written out as follows:\cite{RMP,MFA,B}
%\begin{widetext}
\begin{equation}
\smallskip H_{BG}=-\sum\limits_{i>j}t_{ij}(B_{i}^{\dagger}B_{j}+B_{j}^{\dagger}B_{i})
+\sum\limits_{i>j}V_{ij}N_{i}N_{j}-\mu \sum\limits_{i}N_{i},  \label{Bip}
\end{equation}
%\end{widetext}
where $N_{i}=B_{i}^{\dagger}B_{i}$, $\mu $ is a chemical potential derived from
the condition of fixed full number of bosons $N_{l}=
\sum\limits_{i=1}^{N}\langle N_{i}\rangle $, or concentration $\;n=N_{l}/N\in
[0,1]$. The $t_{ij}$ denotes an effective transfer integral,  $V_{ij}$ is an
intersite interaction between the bosons. \smallskip Here $B^{\dagger}(B)$ are
the Pauli creation (annihilation) operators which are Bose-like commuting for
different sites $[B_{i},B_{j}^{\dagger}]=0,$ for $i\neq j,$ and  for the same site $%
B_{i}^{2}=(B_{i}^{\dagger})^{2}=0$, $[B_{i},B_{i}^{\dagger}]=1-2N_i$, $N_i =
B_{i}^{\dagger}B_{i}$; $N$ is a full number of sites.

The disproportionated phase as well as the lattice hard-core Bose-gas  is
equivalent to a system of spins $s=1/2$  exposed to an external magnetic field in the $z$%
-direction. Indeed, the charge $(e,h)$, or $M^{\mp}$-doublet, that is two
different valence states of $M$-centers,   one might associate with two
possible states   of the {\it charge pseudo-spin} (isospin) $s=\frac{1}{2}$:
$|+\frac{1}{2}\rangle$ and $|-\frac{1}{2}\rangle$ for electron $M^-$ and hole
$M^+$ centers, respectively. Then the effective Hamiltonian can be written as
follows \cite{RMP,MFA,B}
\begin{equation}
H_{BG}=\sum_{i>j}J^{xy}_{ij}(s_{i}^{+}s_{j}^{-}+s_{j}^{+}s_{i}^{-})+\sum\limits_{i>j}
J^{z}_{ij}s_{i}^{z}s_{j}^{z}-h \sum\limits_{i}s_{i}^{z}, \label{spinBG}
\end{equation}
where $J^{xy}_{ij}=2t_{ij}$, $J^{z}_{ij}=V_{ij}$, $h =\mu -\sum\limits_{j
(j\neq i)}V_{ij}$, $s^{-}= \frac{1}{\sqrt{2}}B_ , s^{+}=-\frac{1}{\sqrt{2}}
 B^{\dagger}, s^{z}=-\frac{1}{2}+B_{i}^{\dagger}B_{i}$,
$s^{\pm}=\mp \frac{1}{\sqrt{2}}(s^x \pm \imath s^y)$. Below we make use of
conventional two-sublattice approach.  For the  description of the pseudospin
ordering to be more physically clear one may introduce two vectors, the
ferromagnetic and antiferromagnetic ones:
$$
{\bf m}=\frac{1}{2s}(\langle{\bf s}_1 \rangle +\langle{\bf s}_2 \rangle);\,
{\bf l}=\frac{1}{2s}(\langle{\bf s}_1 \rangle -\langle{\bf s}_2
\rangle);\,\,{\bf m}^2 +{\bf l}^2 =1\, ,
$$
where ${\bf m}\cdot {\bf l}=0$.
The hard-core boson system in a two-sublattice approximation is described by
two diagonal order parameters $l_z ,m_z$ and two complex off-diagonal
 order parameters: $m_{\pm}=\mp \frac{1}{\sqrt{2}}(m_x \pm \imath m_y)$ and
$l_{\pm}=\mp \frac{1}{\sqrt{2}}(l_x \pm \imath l_y)$. The complex superfluid
order parameter $\Psi ({\bf  r})=|\Psi ({\bf  r})|\exp-\imath\phi $ is
determined by the in-plane components of ferromagnetic vector: $ \Psi ({\bf 
r})=\frac{1}{2}\langle (\hat B_1 +\hat B_2 )\rangle
=sm_{-}=sm_{\perp}\exp-\imath\phi $, $m_{\perp}$ being the length of the
in-plane component of ferromagnetic vector. So, for a local condensate density
we get $n_s = s^2 m_{\perp}^2$.  It is of interest to note that in fact all the
conventional uniform $T=0$ states with nonzero $\Psi ({\bf  r})$ imply
simultaneous long-range order both for modulus $|\Psi ({\bf  r})|$ and phase
$\phi$. The in-plane components of antiferromagnetic vector $l_{\pm}$ describe
a staggered off-diagonal order.

  The model exhibits many fascinating quantum phases and phase
transitions. Early investigations predict at $T=0$ charge order (CO), Bose
superfluid (BS) and mixed (BS+CO) supersolid uniform phases with an Ising-type
melting transition (CO-NO) and Kosterlitz-Thouless-type (BS-NO) phase
transitions to a non-ordered normal fluid (NO) in 2D systems.\cite{RMP,MFA,B}

\section{Topological charge fluctuations in model mixed-valence system }
Above we focused on the homogeneous phase states of the mixed-valence system.  
Main short-length scale charge fluctuations in $M^0$ and $M^{\pm}$ systems are
associated with a thermal exciton creation, or annihilation due to a reaction:
$(M^0 - M^0) \leftrightarrow (M^+ - M^- )$. Amongst the long-length scale
charge fluctuations in a model system we would like to address the 
topological defects in quasi-2D systems such as cuprates, in particular, different  bubble-like entities like skyrmions, or another out-of-plane vortices. Namely these one can play the main role in a
nucleation of unconventional charge phases. 

\subsection{Topological defects in $M^0$-phase}
 The most interesting situation concerns the charge fluctuations in
the {\it conceptually simple} monovalent $M^0$-phase which seems to be a
representative of traditional insulating oxides. From the viewpoint of the
${\bf  a},{\bf  b}$-vector
 field formalism  the $M^0$ system resembles, in a sense, the easy-axis
 (anti)ferromagnet. Indeed, the phase is specified by a simple uniform arrangement
  of ${\bf  a}$ and ${\bf  b}$ vectors parallel to $z$-axis:
   ${\bf  a}\parallel {\bf  b}\parallel O_z$, and its energy does not depend on
 the sense of these vectors. The analogy allows us to make use of simple
 physically clear pictures of (anti)ferromagnetic domain structures.

Below we'll  address two types of domain walls in $M^0$-phase. The first would
illustrate the long-scale fluctuation which conserves the mean on-site valence:
in other words, $\langle S_z \rangle =0$. The second would provide the example
of a domain wall with a checkerboard charge ordering in its center.

 First of all, let note that  instead of two parallel  vectors
${\bf  a}$ and ${\bf  b}$ given the normalization condition,  the $M^0$-phase
can be described by a unified vector ${\bf  n}$: $\bf  a=\alpha\bf  n, \bf 
b=\beta\bf  n$, and $\alpha+i\beta= \exp (i\kappa)$, $\kappa\in R$. Hereafter,
we denote $\bf  n=n\{\sin\theta\cos\phi, \, \sin\theta\sin\phi, \,
\cos\theta\}$. Moreover, we may introduce a multitude of phases which differ
only by the orientation of the unit vector ${\bf  n}$, or ${\bf  n}$-phases.
Such phases may be considered to be the solutions of a purely biquadratic
pseudo-spin Hamiltonian \cite{electr}
 \begin{equation}
 {\hat H}_{bq}=-J_2\sum_{i,\eta}\sum_{k\geq
j}^3(\{\hat{S}_k\hat{S}_j\}_i\{\hat{S}_k\hat{S}_j\}_{i+\eta}), \label{bq}
 \end{equation}
which can be obtained from (\ref{su3}) given the only nonzero "exchange"
parameters $J_{kk}$ with $k=4\div 8$. The quantum Hamiltonian can be mapped
onto classical Hamiltonian \cite{electr}
\begin{equation}
 H_{bq}= 2J_2|{\bf n}|^{2}\int d^2{\bf r}\left[\sum_{i=1}^{3}(\bf {\nabla}n_i)^2\right]\, .
\label{hm4}
\end{equation}
The constant zero value of the  mean on-site valence $\langle S_z \rangle =0$
is the common property of the ${\bf  n}$-phases. Moreover,  the mean value  of
all the pseudo-spin components turns into zero: $\langle {\bf  S} \rangle =0$.
Let address the ferromagnetic ${\bf  n}$-phase with $180^o$ domain walls which
separate two actually equivalent $M^0$-domains with opposite direction of
${\bf  n}$ vectors. The picture differs from that of  conventional ferromagnets
where $180^o$ domain wall separates the domains with the opposite direction of
magnetization. In the center of such a {\it BS-type} domain wall we may deal
with a superfluid BS phase ${\bf  n}\perp O_z$.

In 2D systems such as cuprates it appears possible to form skyrmion-like
topological defects like bubbles. \cite{electr} The skyrmion spin texture looks
like  a bubble domain in ferromagnet and consists of a vortex-like arrangement
of the in-plane components of spin with the $z$-component reversed in the
centre of the skyrmion and gradually increasing to match the homogeneous
background at infinity. The spin distribution within a classical Belavin-Polyakov (BP) skyrmion  is
given as follows \cite{BP}
\begin{equation}
\phi=q\varphi +\varphi
_0;\quad\cos\theta=\frac{r^{2q}-\lambda^{2q}}{r^{2q}+\lambda^{2q}}, \label{sk}
\end{equation}
or for the winding number $q=1$
\begin{equation}
n_{x}=\frac{2r\lambda }{r^{2}+\lambda ^{2}}\cos \phi ;\, n_{y}=\frac{2r\lambda
}{r^{2}+\lambda ^{2}}\sin \phi ;\, n_{z}=\frac{r^{2}-\lambda
^{2}}{r^{2}+\lambda ^{2}}.\label{sk1}
\end{equation}
Skyrmions are characterized by the magnitude and sign of its topological
charge, by its size (radius) $\lambda$, and by the global orientation of the
spin, or U(1) order parameter $\varphi _0$. The scale invariance of skyrmionic
solution reflects in that its energy
%$E_q =8\pi
%qJS^2$
is proportional to  winding number and does not depend on  radius,
 and global phase $\varphi _0$.
An interesting example of topological inhomogeneity is provided by a
multi-center BP skyrmion \cite{BP} which energy does not depend
on the position of the centers. The latter are believed to be addressed as  an
additional degree of freedom, or positional order parameter.

The classical Hamiltonian (\ref{hm4}) has skyrmionic solutions, but instead of
the spin distribution in conventional BP skyrmion \cite{BP} we
deal with a zero spin,
 but a non-zero distribution of five spin-quadrupole moments
  $\langle \Lambda ^{(4,5,6,7,8)}\rangle$, or
   $\langle \{S_{i}S_{j}\}\rangle$ which in turn are determined
by the skyrmionic texture  of the $\bf  n$ vector:
  \begin{widetext}
\begin{equation}
\langle (S_x^2 + S_y^{2})\rangle =1+\cos^{2}\theta \,;\langle S_z^2\rangle
=\sin^{2}\theta \,; \langle (S_x^2 - S_y^{2})\rangle =-\sin^{2}\theta
\cos2(q\varphi + \varphi _{0})\,; \label{qq}
\end{equation}
$$
\langle \{S_xS_y\}\rangle =-\sin^{2}\theta \sin2(q\varphi + \varphi _{0});
\langle \{S_xS_z\}\rangle =-\sin2\theta \cos(q\varphi + \varphi _{0}); \langle
\{S_yS_z\}\rangle =-\sin2\theta \sin(q\varphi + \varphi _{0}).
$$
\end{widetext}

Thus we arrive at the ring-shaped distribution of the effective ionicity
$\langle S_z^2\rangle$ with the maximal value of $1$ ($M^{\pm}$-phase) at
$r=\lambda$, and the minimal value of $0$ at the ring center $r=0$ and far
outside $r\rightarrow \infty$ ($M^0$-phase). The "bosonic" off-diagonal complex
order parameter $\langle V_{\pm 2}^2\rangle \propto \langle S_{\pm}^2\rangle
\propto \sin^{2}\theta \exp(\pm 2(q\varphi + \varphi _{0}))$ has a similar
$r$-dependence, while the "fermionic" off-diagonal complex order parameter
$\langle V_{\pm 1}^2\rangle \propto \sin2\theta \exp(\pm (q\varphi + \varphi
_{0}))$ turns into zero both at the ring center $r=0$, far outside
$r\rightarrow \infty$ ($M^0$-phase), and at $r=\lambda$, or everywhere, where
the ionicity has a strictly definite value.

 Despite these  skyrmions are derived from the toy model,  they
yield very instructive information as regards  the probable spin texture of
real solitons with BS-type domain walls and "superconductive" fluctuations in
$M^0$-phase.

 The {\it CO-type} domain walls with a nonzero mean on-site valence given the total
  $\sum_{i}\langle S_z \rangle =0$  can be obtained, in common, in a two-sublattice approximation with
  non-collinear $\bf  a$ and $\bf  b$ vectors.
  Let assume the  $M^0$-phase to be divided onto two sublattices, A and B,
 with  ${\bf  a}_A ={\bf  b}_A ={\bf  a}_B ={\bf  b}_B $ and ${\bf  l}_A ={\bf  l}_B =0$.
Then the CO-type domain wall may be represented by a gradual spatial
non-equivalent rotation of $\bf  a$ and $\bf  b$ vectors in A and B sublattices
providing the nonzero magnitude of $l_z$ components given $l_{zA}=-l_{zB}$ with
its maximum in the center of the domain wall.

\subsection{Topological phase separation in 2D EH Bose liquid away from half-filling}
One of the fundamental hot debated  problems in bosonic physics concerns the
evolution of the charge ordered (CO) ground state of 2D hard-core BH model
(hc-BH)  with a doping away from half-filling.
Numerous model studies steadily confirmed the emergence of "supersolid" phases
with simultaneous diagonal and off-diagonal long range order while Penrose and
Onsager \cite{Penrose} were the first showing as early as 1956 that supersolid
phases do not occur.
The most recent quantum Monte-Carlo  (QMC) simulations
\cite{Batrouni,Hebert,Schmid} found two significant features of the
2D Bose-Hubbard model with a
 screened Coulomb repulsion: the absence of supersolid
phase  at half-filling, and a  growing tendency to phase separation (CO+BS)
upon doping away from half-filling.  Moreover, Batrouni and Scalettar
\cite{Batrouni} studied quantum phase transitions in the ground state
of the 2D hard-core boson Hubbard Hamiltonian and have shown  numerically that,
contrary to the generally held belief, the most commonly discussed
"checkerboard" supersolid is thermodynamically unstable and phase separates
into solid and superfluid phases. The physics of the CO+BS phase separation in
Bose-Hubbard model is associated with a rapid increase of the energy of a
homogeneous CO state with doping away from half-filling due to a large
"pseudo-spin-flip" energy cost.
Hence, it appears to be
energetically more favorable to "extract" extra bosons (holes) from the CO
state and arrange them into finite clusters with a relatively small number of
particles. Such a droplet scenario is believed  to minimize the long-range
Coulomb repulsion.

Magnetic analogy allows us to make unambiguous predictions as
regards the doping of BH system away from half-filling. Indeed,
the boson/hole doping of checkerboard CO phase corresponds to the magnetization
of an antiferromagnet in $z$-direction. In the uniform easy-axis $l_z$-phase of
anisotropic antiferromagnet the local spin-flip energy cost is rather big. In
other words, the energy cost for boson/hole doping into  checkerboard CO phase
appears to be big one due to a large contribution of boson-boson repulsion.
However, the  magnetization of the  anisotropic antiferromagnet in
an easy axis direction may proceed as a first order phase transition
with a ``topological phase separation'' due to the existence of
antiphase domains. The antiphase domain walls
provide the natural nucleation  centers for a spin-flop phase having
enhanced magnetic susceptibility as compared with small if any
longitudinal susceptibility thus  providing the advantage of the
field energy. Namely domain walls  would specify the inhomogeneous
magnetization pattern for such an  anisotropic  easy-axis
antiferromagnet in relatively weak external magnetic field. As
concerns the domain type in quasi-two-dimensional antiferromagnet
one should emphasize the specific role played by the cylindrical, or
bubble domains which have finite energy and size. These topological
solitons have the vortex-like in-plane spin structure and resemble
 classical, or Belavin-Polyakov  skyrmions.\cite{BP}  Although  some questions were
not completely clarified and remain open until now,
 the classical and quantum  skyrmion-like topological defects are
believed to be a genuine element of essential physics both of ferro-
and antiferromagnetic 2D easy-axis systems.
The magnetic analogy seems to be a little bit naive, however, it catches the
essential physics of doping the hc-BH system.
      As regards the checkerboard CO phase of such a  system, namely a  finite
energy  skyrmion-like  bubble domain \cite{bubble,bubble1}     seems to be the  most
preferable candidate for the  domain with antiphase domain wall providing the
natural reservoir for extra bosons.
The skyrmion spin texture
consists of a vortex-like
arrangement of the in-plane components of ferromagnetic ${\bf m}$ vector  with
the $l_z$-component reversed
in the centre of the skyrmion and gradually increasing to match the homogeneous
background at infinity. 

Recently \cite{bubble1} it was shown that the doping, or deviation from half-filling in 2D EH Bose liquid is accompanied by the formation of multi-center topological defect such as charge order (CO) bubble domain(s) with Bose superfluid (BS) and extra bosons both localized in domain wall(s), or a  topological CO+BS  phase separation, rather than an uniform mixed CO+BS supersolid phase.  

 The most probable possibility is that every bubble accumulates one or two
particles. Then, the number of such entities in a multigranular texture
nucleated with doping  has to  nearly linearly depend  on the doping.
   Generally speaking, each individual bubble may be characterized by its position,  nanoscale size, and the
orientation of U(1) degree of freedom. In contrast with the uniform states the phase of
the superfluid order parameter for a bubble is assumed to be unordered. 
In the long-wavelength limit the off-diagonal ordering can be described by an
effective Hamiltonian in terms of  U(1) (phase) degree of freedom associated
with each bubble. Such a Hamiltonian
 contains a repulsive, long-range Coulomb part and a
short-range contribution related to the phase degree of freedom. The
latter term can be written out in the standard for the $XY$ model form of a
so-called Josephson coupling \cite{Kivelson,bubble,CEF}
\begin{equation}
H_J = -\sum_{\langle i,j\rangle}J_{ij}\cos(\varphi _{i}-\varphi _{j}),
\end{equation}
where $\varphi _{i},\varphi _{j}$ are global phases for micrograins centered at
points $i,j$, respectively, $J_{ij}$ Josephson coupling parameter. Namely the
Josephson coupling gives rise to the long-range ordering of the phase of the
superfluid order parameter in such a multi-center texture. Such a Hamiltonian
represents a starting point for the analysis of disordered superconductors,
granular superconductivity, insulator-superconductor transition with $\langle
i,j\rangle$ array of superconducting islands with phases $\varphi _{i},\varphi
_{j}$.
To account for Coulomb interaction and allow for quantum corrections we should
introduce into effective Hamiltonian  the charging energy \cite{Kivelson}
$$
H_{ch}=-\frac{1}{2}q^2 \sum_{i,j}n_{i}(C^{-1})_{ij}n_{j}\, ,
$$
where $n_{i}$ is a  number operator for particles bound in $i$-th micrograin;
it
is quantum-mechanically conjugated to $\varphi$: $n_{i}=-i \partial /\partial
\varphi
_{i}$, $(C^{-1})_{ij}$ stands for  the capacitance matrix, $q$ for a particle
charge.

 Such a system appears to reveal a tremendously rich quantum-critical
structure.\cite{Green,Timm} In the absence of disorder, the
$T=0$ phase diagram of the multi-bubble system implies either triangular, or
square crystalline arrangements  with possible melting transition to a  liquid.
 It should be noted that analogy with charged $2D$ Coulomb gas
implies the Wigner crystallization of  multi-bubble system with Wigner
crystal (WC) to Wigner liquid melting transition, respectively. Naturally, that
the
additional degrees of freedom for a bubble provide a richer physics of
such lattices. For a system  to be an insulator, disorder is required, which
pins the multigranular system
and also causes the crystalline order to have a finite correlation length.
Traditional approach to a Wigner crystallization implies the formation of a WC
for densities lower than a critical density, when the Coulomb energy dominates
over the kinetic energy. The effect of quantum fluctuations leads to a
(quantum) melting of the solid at high densities, or at a critical lattice
spacing. The critical properties of a two-dimensional lattice without any
internal degree of freedom are successfully described  applying the BKT (Berezinsky-Kosterlitz-Thowless) theory
to dislocations and disclinations of the lattice, and proceeds in two steps.
The first implies the transition to a liquid-crystal phase with a short-range
translational order, the second does the transition to isotropic liquid.
 For such a system provided the bubble positions  fixed at all temperatures, the long-wave-length physics would be
described by an (anti)ferromagnetic $XY$ model with expectable BKT transition and
gapless $XY$ spin-wave mode.

The low temperature physics in a multi-bubble system is  governed by an
interplay of two BKT transitions,  for the U(1) phase  and positional degrees
of freedom, respectively.\cite{Timm}
Dislocations  lead to a mismatch in the U(1)
degree of freedom, which makes the dislocations bind fractional vortices and
lead to a coupling of translational and phase excitations. Both BKT temperatures
either coincide (square lattice) or the melting one is higher (triangular
lattice).\cite{Timm}

 Quantum fluctuations can substantially affect these
results. Quantum melting can destroy U(1) order at sufficiently
low densities where the Josephson coupling becomes exponentially small. Similar
situation is expected to take place in the vicinity of
structural transitions in a multigranular crystal. With increasing the
micrograin density the quantum effects  result in a significant lowering of the melting
temperature as compared with classical square-root dependence.
The resulting melting temperature can reveal  an oscilating behavior as a
function of particle density with zeros at the critical (magic) densities
associated with structural phase transitions. 

In terms of our model, the positional order corresponds to an incommensurate
charge density wave, while the U(1) order does to a superconductivity. In other
words, we arrive at a subtle interplay between two orders. The superconducting
state evolves from a charge order with $T_C \leq T_m$, where $T_m$ is the
temperature of a melting transition which could be termed as a temperature of
the opening of the insulating gap (pseudo-gap!?).

The normal modes of a dilute  multi-bubble system include the pseudo-spin waves
propagating in-between the bubbles, the positional fluctuations, or
quasi-phonon modes,  which are gapless in  a pure system, but gapped  when
the lattice is pinned, and, finally,  fluctuations in the U(1) order parameter.

    The orientational fluctuations of the multi-bubble system are governed by the gapless
   $XY$ model.\cite{Green} The relevant model description is most familiar as
an effective   theory of the Josephson junction array. An important feature of
the model is that it displays a quantum-critical point.

The low-energy collective excitations of a multi-bubble
liquid includes an usual longitudinal acoustic phonon-like branch.
The liquid crystal phases differ from the isotropic liquid in that they have
massive topological excitations, {\it i.e.}, the disclinations.
One should note that the liquids do not support transverse modes, these could
survive in a liquid state only as overdamped modes.  So that it is reasonable
to assume that solidification of the bubble lattice would be accompanied by a
stabilization of transverse phonon-like modes with its sharpening below melting
transition.
In other words, an instability of transverse phonon-like modes signals the
onset of melting.
The phonon-like modes in the bubble crystal have much in common with usual phonon
modes, however, due to electronic nature these can hardly  be detected if any
by inelastic neutron scattering.

A generic property of the positionally ordered bubble configuration is the
sliding mode which is usually pinned by the disorder. The depinning of sliding
mode(s) can be detected in a low-frequency and low-temperature optical
response.

\section{Implications for strongly correlated oxides}
In this Section we suggest some speculations around an unconventional scenario
of the essential physics of cuprates and manganites that implies their
instability with regard to the self-trapping of charge transfer excitons and the
formation of electron-hole Bose liquid.

\subsection{Cuprates}
 The origin of high-T$_c$
superconductivity is presently still a matter of great controversy.
 The unconventional behavior of
cuprates strongly differs from that of ordinary metals and merely resembles
that of doped semiconductor. Moreover, the history of high T$_c$'s itself shows
that we deal with a transformation of particularly insulating state in which
the electron correlations govern the essential physics.

The copper oxides start out life as insulators in contrast with BCS
superconductors being conventional metals. It is impossible to understand the
behavior of the doped cuprates and, in particular, the origin of HTSC unless
the nature of the doped-insulating state is incorporated into the theory. The
problem of a doped insulator is sure much more complicated than it is implied
in   oversimplified approaches such as an effective $t$-$J$-model when the
situation resembles that of ``throwing the baby out of the bathwater''.

 In  a case of cuprates we deal with systems which conventional ground state seems
 to be unstable with regard to the transformation into a new phase state with
a variety of unusual properties.

In our view, the essential physics of the  doped cuprates, as well as other
strongly correlated oxides, is driven by a self-trapping of the CT excitons,
both one-center, and two-center. \cite{CT,CT1} Such excitons are the result
of self-consistent charge transfer and lattice distortion with the appearance
of the ``negative-$U$'' effect.\cite{Shluger,Vikhnin} The  two-center excitons
are associated with CT transitions between two CuO$_4$ plaquettes, and may be
considered as quanta of the disproportionation reaction
$$
\mbox{CuO}_{4}^{6-}+\mbox{CuO}_{4}^{6-}\rightarrow \mbox{CuO}_{4}^{7-}+
\mbox{CuO}_{4}^{5-}
$$
with the creation of electron  CuO$_{4}^{7-}$ and hole CuO$_{4}^{5-}$ centers.
Thus, three types of CuO$_4$ centers  CuO$_{4}^{5,6,7-}$ should be considered
on equal footing.
 In this connection we would
like to draw special attention to the lattice polarization and relaxation
effects that are of primary importance both for the formation of CT exciton
itself and its self-trapping. It should be noted that the photo-excited
electron-hole pairs or excitons are stabilized into self-trapped excitons (STE)
accompanied with lattice relaxation within several pico-seconds.\cite{Toyozawa}

In contrast with BaBiO$_3$ system where we deal with a  spontaneous generation
of self-trapped CT excitons  in the ground state,   the parent insulating
cuprates are believed to be near excitonic instability when
 the self-trapped CT excitons  form the candidate relaxed excited states to struggle with the
conventional ground state. \cite{Toyozawa} In other words, the lattice relaxed
CT excited state should be treated on an equal footing with the ground state.
If the interaction between STE were attractive and so large that the cohesive
energy $W_1$ per one STE exceeds the energy $E_R$ of one STE , the STE's and/or
its clusters will be spontaneously generated everywhere without any optical
excitation, and be condensed to form a new electronic state on a new lattice
structure. \cite{Toyozawa}
 
The minimal energy cost of the optically excited disproportionation or
electron-hole formation in insulating cuprates is $2.0\div  2.5$ eV.
\cite{CT,CT1} Interestingly, that this relatively small value of the optical
gap was nevertheless used by Goodenough \cite{Good} as argument against the
``negative-U'' disproportionation reaction 2Cu(II) = Cu(III) + Cu(I), or more
correctly 2[CuO$_{4}^{6-}$]=[CuO$_{4}^{5-}$]+[CuO$_{4}^{7-}$] in cuprates.
However, the question arises, what is the energy cost for the thermal
excitation of such a disproportionation? In other words, what is the energy of
self-trapped two-center CT exciton? It is this quantity rather than its optical
counterpart defines the activation energy for such a reaction.   The question
is of primary importance for the self-trapping scenario. The answer implies
first of all the knowledge of electronic and ionic polarization energies for
electron and hole. The polarization effects with its typical energy scale of
the order of several eV appear to be of primary importance when one considers
different  charge fluctuations in insulators.

Several general theories of self-trapping have been proposed starting from
works by Landau, Pekar and Toyozawa, however, the stages of the ST process and
detailed atomistic and electronic structure of ST-excitons and ST-holes are
still unclear even in rather simple and well-studied systems such as
alkali-metal halides.

As regards the STE in cuprates, we have some straightforward experimental
indications.  A key characteristic of the STE is its luminescence: STE are
short-lived luminescent states of excited crystals. The observation of
photoluminescence (PL) near $2.0\div 2.4$ eV in La$_2$CuO$_4$,
\cite{Salamon,Ginder} near $1.3$ and $2.4$ eV in YBa$_2$Cu$_3$O$_{6}$,
\cite{Salamon,Denisov,Eremenko} near $1.78,\,1.95,\,2.06$ eV in
PrBa$_2$Cu$_3$O$_{6}$ \cite{Salamon} is a direct evidence of strongly localized
long-lived states related to self-trapped excitons or their derivatives. The
near-infrared photoluminescence   was observed in many insulating cuprates.
\cite{Denisov,Salamon,Sugai} To the best of our knowledge the most detailed PL
study was performed by Denisov {\it et al}. \cite{Denisov} in
YBa$_2$Cu$_3$O$_{6+x}$ in the spectral range $0.7\div 1.4$ eV. The
low-temperature PL in YBa$_2$Cu$_3$O$_6$ consists of three peaks at $0.87$,
$1.07$, and $\sim 1.4$ eV, respectively. The PL intensity is much stronger at
small doping level. Moreover, the doping induced PL suppression manifests
itself  more strongly for the low-energy  than for the high-energy PL peaks. At
$x=0.15$ only the high-energy peak located at $1.28$ eV ($T=10 \,K$) survives
that allows us to assume that the STE decay becomes more effective with doping.
The high-energy PL peak red-shifts with the lowering the temperature, and its
intensity decreases.

All these features  can be consistently explained in frames of the STE nature
of PL. Different PL peak can be assigned both to different STE and its clusters
pointing to the multistage character of the luminescence.

Cuprates are believed to be unconventional systems which are unstable with
regard to a self-trapping  of the low-energy charge transfer excitons with a
nucleation of electron-hole (EH) droplets being actually the system of coupled
electron CuO$_{4}^{7-}$ and hole CuO$_{4}^{5-}$ centers having been glued in
lattice due to a strong electron-lattice polarization effects. Phase transition
to novel hypothetically metallic state  could be realized due to a
  mechanism familiar to semiconductors with filled bands such as Ge and Si where
  given  certain conditions one observes
   a formation of metallic EH-liquid as a result of the exciton decay.
   \cite{Rice}
 The  system of strongly correlated electron CuO$_{4}^{7-}$ and hole
CuO$_{4}^{5-}$ centers  appears to be equivalent to an electron-hole
Bose-liquid (EHBL) in contrast with the electron-hole  Fermi-liquid in
conventional semiconductors. A simple model description of such a liquid
implies
 a system of local singlet bosons with a charge of $q=2e$ moving in a lattice
  formed by hole centers.
Local boson in our scenario represents the electron counterpart of Zhang-Rice
singlet, or two-electron configuration $b_{1g}^{2}{}^{1}A_{1g}$.
  Naturally, that conventional electron CuO$_{4}^{7-}$ center represents a
  relaxed state of composite system: "hole CuO$_{4}^{5-}$ center plus local
  singlet boson", while the "non-retarded" scenario of a novel phase
is assumed to incorporate the unconventional states of electron CuO$_{4}^{7-}$
center up to its orbital degeneracy.

Thus we can introduce the concept of insulator-to-EHBL transition as the
spontaneous condensation of self-trapped excitons and its clusters. However, in
cuprates we deal with
 the electron/hole injection to the insulating
parent phase  due to a nonisovalent substitution as in
La$_{2-x}$Sr$_x$CuO$_{4}$, Nd$_{2-x}$Ce$_x$CuO$_{4}$, or change in oxygen
stoihiometry as in YBa$_2$Cu$_3$O$_{6+x}$, La$_{2}$CuO$_{4-\delta}$,
La$_{2}$Cu$_{1-x}$Li$_x$O$_{4}$. Such a substitution provokes the nucleation of
EH droplets and shifts the phase equilibrium from the insulating state to the
unconventional electron-hole Bose liquid, or, in other words,  induces the
insulator-to-EHBL phase transition. Hence, the formation of  EHBL in cuprates
can be considered as the first order phase transition. The doping in cuprates
gradually shifts the EHBL state away from half-filing.

It is clear that the EHBL scenario makes the doped cuprates the objects of
$bosonic$ physics. There are numerous experimental evidence that support the
bosonic scenario for doped cuprates.\cite{ASA} In this connection, we would
like to draw attention to the little known results of comparative
high-temperature studies of thermoelectric power and conductivity which
unambiguously reveal the charge carriers with $q=2e$, or two-electron(hole)
transport.\cite{Victor} The well-known relation $\frac{\partial
\alpha}{\partial \ln \sigma}=const=-\frac{k}{q}$ with $|q|=2|e|$ is fulfilled
with high accuracy in the limit of high temperatures ($\sim 700\div 1000 K$)
for different cuprates (YBa$_{2}$Cu$_{3}$O$_{6+x}$,
La$_{3}$Ba$_{3}$Cu$_{6}$O$_{14+x}$,
(Nd$_{2/3}$Ce$_{1/3}$)$_{4}$(Ba$_{2/3}$Nd$_{1/3}$)$_{4}$Cu$_{6}$O$_{16+x}$).

 \subsection{Manganites}
 Parent manganites  such
as LaMnO$_3$ are antiferromagnetic insulators with the charge transfer
 gap.   Fundamental absorption band in parent manganites is formed both by the
   intracenter O2p-Mn3d transfer \cite{LMnO3}  and by the small intercenter charge transfer
    excitons,\cite{Khaliullin} which  in terms of chemical notions represent somewhat like the disproportionation
     quanta with a rather low threshold of the order of 3 eV, resulting in a formation
     of electron MnO$_6^{10-}$ and hole
 MnO$_6^{8-}$ centers.  The CT excitons in LaMnO$_3$  prone to a  self-trapping \cite{Kovaleva}
 and may be considered to be well defined entities only at
 small content, whereas at large densities their coupling is screened and their overlap becomes
  so considerable  that they loose individuality, become unstable with regard to the decay
  (the dissociation) to  electron  and hole  centers, and we come to a system of electrons and
  holes, which form an  EH Bose liquid. 

An instability of parent manganite LaMnO$_3$ with regard to the overall
disproportionation such as
\begin{equation}
\mbox{MnO}_{6}^{9-}+ \mbox{MnO}_{6}^{9-} \rightarrow \mbox{MnO}_{6}^{10-}+
\mbox{MnO}_{6}^{8-} \label{dispro}
\end{equation}
 was strikingly demonstrated recently by Zhou and Goodenough. \cite{Zhou} The transport
 (thermoelectric power and resistivity) and magnetic (susceptibility) measurements showed that
  LaMnO$_3$ above the cooperative Jahn-Teller orbital-ordering temperature $T_{JT}\approx 750$ K
   transforms into charge-disproportionated paramagnetic phase with $\mu _{eff}=5.22 \mu _{B}$
    and cooperative charge transfer of many heavy vibronic charge carriers. It
    seems rather obvious that with the lowering the temperature we arrive at a
    system with the well developed fluctuations of the EH Bose-liquid phase.
Strong variation of the LaMnO$_3$ Raman spectra with increasing laser power
\cite{Raman} could be related to the photo-induced nucleation and  the volume
expansion of the EH Bose-liquid, especially, as at a rather big excitation
wavelength $\lambda = 514.5$ nm, at $\lambda = 632.8$
 nm the absorption is considerably stronger in the domains of novel phase than in
 parent lattice.

Effective nucleation of the EH Bose-liquid in manganites could be provoked by a
non-isovalent substitution since this strongly polarizable or even metallic
phase in contrast with parent insulating phase  provides an effective screening
of charge inhomogeneity. Indeed, in thermoelectric power (TEP) experiments with
doped manganites such as La$_{1-x}$Sr$_x$MnO$_3$ Hundley and Neumeier
\cite{TEP} have found that more hole-like charge carriers or alternatively
fewer accessible Mn sites are present than expected for the value $x$. They
suggest a charge disproportionation model based on the instability of
 Mn$^{3+}$-Mn$^{3+}$ relative to that of Mn$^{4+}$-Mn$^{2+}$. This transformation provides excellent
  agreement with doping-dependent trends exhibited by both TEP and resistivity.

 A simplified "chemical" approach to an EH Bose-liquid as to a
disproportionated phase \cite{Ionov} naively implies an occurrence of Mn$^{4+}$
and Mn$^{2+}$ ions. However, such an approach is very far from reality. Indeed,
the electron MnO$_6^{10-}$ and  hole MnO$_6^{8-}$ centers are already the mixed
valence centers,\cite{Mn-ehl} as in the former the $Mn$ valence resonates between $+2$ and
$+1$, and in the latter does between $+4$ and $+3$.

In this connection, one should note that in a sense disproportionation reaction
(\ref{dispro}) has several purely ionic counterparts, the two rather simple

Mn$^{3+}$-O$^{2-}$-Mn$^{3+} \rightarrow$ Mn$^{2+}$-O$^{2-}$-Mn$^{4+}$,

and

Mn$^{3+}$-O$^{2-}$-Mn$^{3+} \rightarrow$ Mn$^{2+}$-O$^{1-}$-Mn$^{3+}$,

and, finally, one rather complicated

O$^{2-}$-Mn$^{3+}$-O$^{2-} \rightarrow $ O$^{1-}$-Mn$^{1+}$-O$^{1-}$.

Thus, the disproportionation (\ref{dispro}) threshold energy has to be
maximally close to the CT energy parameter $\Delta _{pd}$. Moreover, namely
this is seemingly to be one of the main parameters governing
 the nucleation of EH Bose-liquid in oxides.

So far, there has been no systematic exploration of exact valence and spin
state of Mn in  these systems. Park {\it et al}. \cite{Park} attempted to
support the Mn$^{3+}$/Mn$^{4+}$ model, based on the Mn 2p x-ray photoelectron
spectroscopy (XPES) and O1s absorption. However, the significant discrepancy
between the weighted Mn$^{3+}$/Mn$^{4+}$ spectrum and the experimental one for
given
 $x$ suggests a more complex doping effect. Subias et al.\cite{Subias} examined the valence
 state of Mn utilizing Mn $K$-edge x-ray  absorption near edge spectra (XANES), however,
  a large discrepancy is found between experimental spectra given intermediate doping and
  appropriate superposition of the end members.

The valence state of Mn in Ca-doped LaMnO$_3$ was studied by high-resolution Mn
$K\beta $ emission spectroscopy by Tyson {\it et al.} \cite{Tyson}. No evidence
for Mn$^{2+}$ was observed at any $x$ values seemingly ruling out proposals
regarding  the Mn$^{3+}$ disproportionation. However, this conclusion seems to
be absolutely unreasonable. Indeed, electron center
 MnO$_{6}^{10-}$ can be found in two configuration with formal Mn valence Mn$^{2+}$ and
 Mn$^{1+}$ (not simply Mn$^{2+}$!), respectively. In its turn, the hole center MnO$_{6}^{8-}$
 can be found in two configurations with formal Mn valence Mn$^{4+}$ and Mn$^{3+}$ (not simply
  Mn$^{4+}$ !), respectively. So, within the model the Mn $K\beta $ emission spectrum for
  Ca-doped LaMn$O_3$ has to be a superposition of appropriately weighted Mn$^{1+}$, Mn$^{2+}$,
   Mn$^{3+}$, Mn$^{4+}$ contributions (not simply Mn$^{4+}$ and Mn$^{3+}$, as one assumes in
  Ref. \onlinecite{Tyson}). Unfortunately, we do not know the Mn $K\beta $ emission spectra for the oxide
   compounds  with Mn$^{1+}$ ions, however a close inspection of the Mn $K\beta $ emission
   spectra for the series of  Mn-oxide compounds with Mn valence varying from $2+$ to $7+$
   (Fig.2 of the cited paper) allows to uncover a rather clear dependence on valence, and
   indicates a possibility to explain the experimental spectrum for Ca-doped LaMnO$_3$
    as a superposition of appropriately weighted Mn$^{1+}$, Mn$^{2+}$, Mn$^{3+}$, Mn$^{4+}$
    contributions. It should be noted that an "arrested" Mn-valence response to the doping
in the $x<0.3$ range founded in Ref. \onlinecite{Tyson} is also consistent with
the creation of predominantly oxygen holes.

This set of conflicting data together with a number of additional data
\cite{Croft} suggests the need for an in-depth exploration of the Mn valence
problem  in this perovskite system. However, one might say, the doped
manganites are not only systems with   mixed valence, but systems with
$indefinite$ valence, where we cannot, strictly speaking,  unambiguously
distinguish Mn species with either distinct valence state.

\section{Conclusions}
 
 We have developed a model approach to describe  charge fluctuations and different charge phases in strongly correlated 3d oxides. In frames of $S=1$ pseudo-spin formalism different phase states of the
system of the metal-oxide $M$ centers with three different valent state $M^{0,\pm}$ are
considered on the equal footing. Simple uniform mean-field   phases include an insulating monovalent $M^0$-phase, mixed-valence binary (disproportionated) $M^{\pm}$-phase, and mixed-valence ternary (``under-disproportionated'') $M^{0,\pm}$-phase. We consider two first  phases in more details focusing on  the  problem of electron/hole states and different types of excitons in $M^0$-phase and formation of electron-hole Bose liquid in $M^{\pm}$-phase. 

Our consideration was focused mainly on a number of issues seemingly being of primary importance for the various  strongly correlated oxides such as cuprates, manganites, bismuthates, and other systems with
 CT  instability and/or mixed valence. These includes two types of single particle correlated hopping and the two-particle hopping, CT excitons, electron-lattice polarization effects which are shown to be crucial for the stabilization of either phase, topological charge fluctuations, nucleation of droplets of the electron-hole Bose liquid and phase separation effect. We  emphasize an important role of self-trapped CT excitons in typical Mott-Hubbard insulators as candidate  "relaxed excited states" to struggle for stability with ground state and natural nucleation centers for unconventional electron-hole Bose liquid which phase state include the superfluid.
 
 Pseudo-spin formalism  has appeared to be very efficient to reveal and describe different aspects of essential physics for mixed-valence system. We show that the coherent states provide the optimal way both to a correct mean-field approximation  and respective continuous models to describe the pseudo-spin system including  different  topological charge fluctuations, in particular, like domain walls or bubble domains in antiferromagnets. 
 All the insulating systems such as $M^0$-phase may be subdivided to two classes: stable and unstable ones with regard to the formation of self-trapped CT excitons. The latter systems appear to be unstable with regard the formation of CT exciton clusters, or droplets of the electron-hole Bose liquid.  
The model approach suggested is believed to provide a conceptual framework for an in-depth understanding  of physics of
 strongly correlated oxides such as cuprates, manganites, bismuthates, and other systems with
 charge transfer excitonic instability and/or mixed valence.  We shortly discuss an unconventional scenario
of the essential physics of cuprates and manganites that implies their instability with regard to the self-trapping of charge transfer excitons and the formation of electron-hole Bose liquid.

Author acknowledges the stimulating discussions with V. Vikhnin, A.V. Mitin, S.-L. Drechsler, T. Mishonov, R. Hayn, I. Eremin, M. Eremin, Yu. Panov, V.L. Kozhevnikov and support by  SMWK Grant, INTAS Grant No. 01-0654, CRDF Grant No. REC-005,  RFBR Grant No. 04-02-96077.

\end{document}